\documentclass[a4paper,11pt]{article}
\pdfoutput=1     
\usepackage{jheppub}  
\usepackage[T1]{fontenc}

\def\U{\mathrm{u}} \def\V{\mathrm{v}}

 \def\[{\left[} \def\]{\right]}
 \def\K{ \mathbb{K}} \def\E{ \mathbb{E}}

\def\arctanh{\mathrm{arctanh}}    
   \usepackage{ifpdf}
\usepackage{ae} \usepackage[T1]{fontenc}
\usepackage[ansinew]{inputenc} \usepackage{mathrsfs}
\usepackage{amsmath} \usepackage{amssymb} \pdfoutput=1
 \newcommand\re[1]{({\ref{#1}})}
\def\be{\begin{eqnarray} } \def\ee{\end{eqnarray}}
\def\bee{\be\begin{aligned}} \def\eee{\end{aligned} \ee }
  
 \def\IC{{\mathbb{C}}} 
 \def\no{\nonumber} \def\la{\label}  \def\({\left(} \def\){\right)} \def\<{\langle}
\def\>{\rangle } \def\[{\left[} \def\]{\right]} \def\hf{
{\textstyle{1\over 2}} }  
      
   \def\a{\alpha}
\def\b{\beta}   
\def\s{\sigma} \def\t{\tau}  \def\th{\theta}
  
\def\Tr{{\rm Tr}}  \def\Xint#1{\mathchoice
   {\XXint\displaystyle\textstyle{#1}}%
   {\XXint\textstyle\scriptstyle{#1}}%
   {\XXint\scriptstyle\scriptscriptstyle{#1}}%
   {\XXint\scriptscriptstyle\scriptscriptstyle{#1}}%
\!\int} \def\XXint#1#2#3{{\setbox0=\hbox{$#1{#2#3}{\int}$}
\vcenter{\hbox{$#2#3$}}\kern-.5\wd0}} 
\def\dashint{\Xint-} \usepackage{bm} 
 \def\ee{\end{eqnarray}}

       \def\H
{H }

\author[a,b]{ I. Kostov}

\affiliation[a]{Universit\'e Paris-Saclay, CNRS, CEA, Institut de
physique th\'eorique
\\
 91191 Gif-sur-Yvette, France } \affiliation[b]{ Department of
 Physics, Federal University of Esp\'irito Santo, \\
 29075-900, Vit\'oria, Brazil 
}

\emailAdd{ivan.kostov@ipht.fr}

  \abstract{ Basso-Dixon integrals evaluate rectangular fishnets --
  Feynman graphs with massless scalar propagators which form a
  $m\times n$ rectangular grid -- which arise in certain one-trace
  four-point correlators in the `fishnet' limit of ${\mathcal{N}}=4$
  SYM. Recently, Basso {\it et al} explored the thermodynamical limit
  $m\to\infty$ with fixed aspect ratio $n/m$ of a rectangular fishnet
  and showed that in general the dependence on the coordinates of the
  four operators is erased, but it reappears in a scaling limit with
  two of the operators getting close in a controlled way.  In this
  note I investigate the most general double scaling limit which
  describes the thermodynamics when one of two pairs of operators
  become nearly light-like.  In this double scaling limit, the
  rectangular fishnet depends on both coordinate cross ratios.  I show
  that all singular limits of the fishnet can be attained within the
  double scaling limit, including the null limit with the four points
  approaching the cusps of a null square.  A direct evaluation of the
  fishnet in the null limit is presented any $m$ and $n$.  }

\title{\boldmath  Light-cone  limits  of large rectangular  fishnets}

\begin{document} 

\maketitle
\flushbottom

\section{ Introduction}
\label{sec:intro}

It has been known since decades that a series of Feynman graphs with
massless propagators, the so called ladder graphs or simply ladders,
can be evaluated explicitly \cite{USSYUKINA1993363}.  In the last
years it became clear that a vast family of planar Feynman graphs
having a regular structure, the so called fishnet graphs, share this
property.  The term fishnet planar graphs was introduced by by A.
Zamolodchikov \cite{Zamolodchikov:1980mb}, who pointed out that they
can be studied by two-dimensional integrability methods.  The
breakthrough was the invention of the `fishnet conformal field theory'
formulated as a certain projection of the integrable ${\mathcal{N}}=4$ SYM
\cite{Gurdogan:2015csr,Caetano:2016ydc,Grabner:2017pgm}, which opened
the possibility to study fishnet Feynman integrals by adapting the
integrability methods developed in ${\mathcal{N}}=4$ SYM. Later the
integrability of the fishnet CFT was established also in the spirit of
the original paper \cite{Zamolodchikov:1980mb} utilising the regular
iterative structure of the fishnet graphs \cite{Gromov:2017cja}. 

Since a fishnet Feynman graph can be interpreted as a (non-compact)
lattice model with nearest neighbour interaction, there are hopes that
its large fishnets is described by some effective $\sigma$-model.  In
fact, the fishnet graphs with gaussian propagators have been first
considered as a possible discretisation of the world sheet of a
string, but such rigid discretisation does not respect the symmetry of
the string path integral.
 
 Remarkably, conformal fishnet integrals do have holographic
 interpretation in case of periodic boundary conditions
 \cite{Basso:2018agi,Gromov:2019bsj,
 Gromov:2019aku,Gromov:2019jfh,Basso:2019xay}.  There is strong
 evidence \cite{Basso:2018agi} that large periodic fishnet diagrams
 correspond to an integrable lattice regularisation of the
 two-dimensional AdS$_5$ non-linear sigma model.

It is believed that a holographic interpretation should exist also for
large open fishnet graphs which are typically associated with
correlation functions of products of matrix operators along a trace.
The open fishnets possess nice integrability properties including
Yangian symmetry \cite{Drummond:Yangian,Frassek:2013xza,Chicherin:2017frs,Chicherin:2022nqq}, but for the moment it is not clear
how to use them to explore their continuum limit in general.
Gathering `experimental' results concerning the thermodynamical limit
of open fishnets might give us some intuition about its $\sigma$-model
description.
  
 In a recent work \cite{Basso:2021omx}, the continuum limit was
 computed in for the simplest open fishnet graphs, introduced
 previously by Basso and Dixon \cite{Basso:2017jwq}, which are
 described briefly below.  Basso-Dixon fishnets correspond to the
 four-point correlators
 \bee \la{fourptcor} G_{m,n}( x_1,x_2,x_3,x_4 ) &= \langle \textrm{Tr}
 \{ \phi_2^n(x_1) \phi_1^m(x_2) \phi_2^{\dagger n}(x_3)
 \phi_1^{\dagger m} (x_4) \} \rangle\, , \eee
in a theory of two $N_c \times N_c $ complex matrix fields $\phi_{1 }$
and $\phi_2$ with chiral quartic interaction $\sim g^2
\Tr[\phi_1\phi_2\phi_1^\dag\phi_2^\dag]$.  The perturbative series for
the correlator $G_{m,n}$ consists of a single Feynman graph
representing regular square lattices of size $m\times n$ with the
external legs on each side attached to four distinct points in the
Minkowski space.  Up to a standard factor, $G_{m,n}$ depends on the
positions of the operators through the two conformal cross ratios,
\bee
\la{openfish}
 G_{m,n}( x_1,x_2,x_3,x_4 ) & 
  \   = \ \, 
\begin{gathered}
\includegraphics[height=2.5cm]{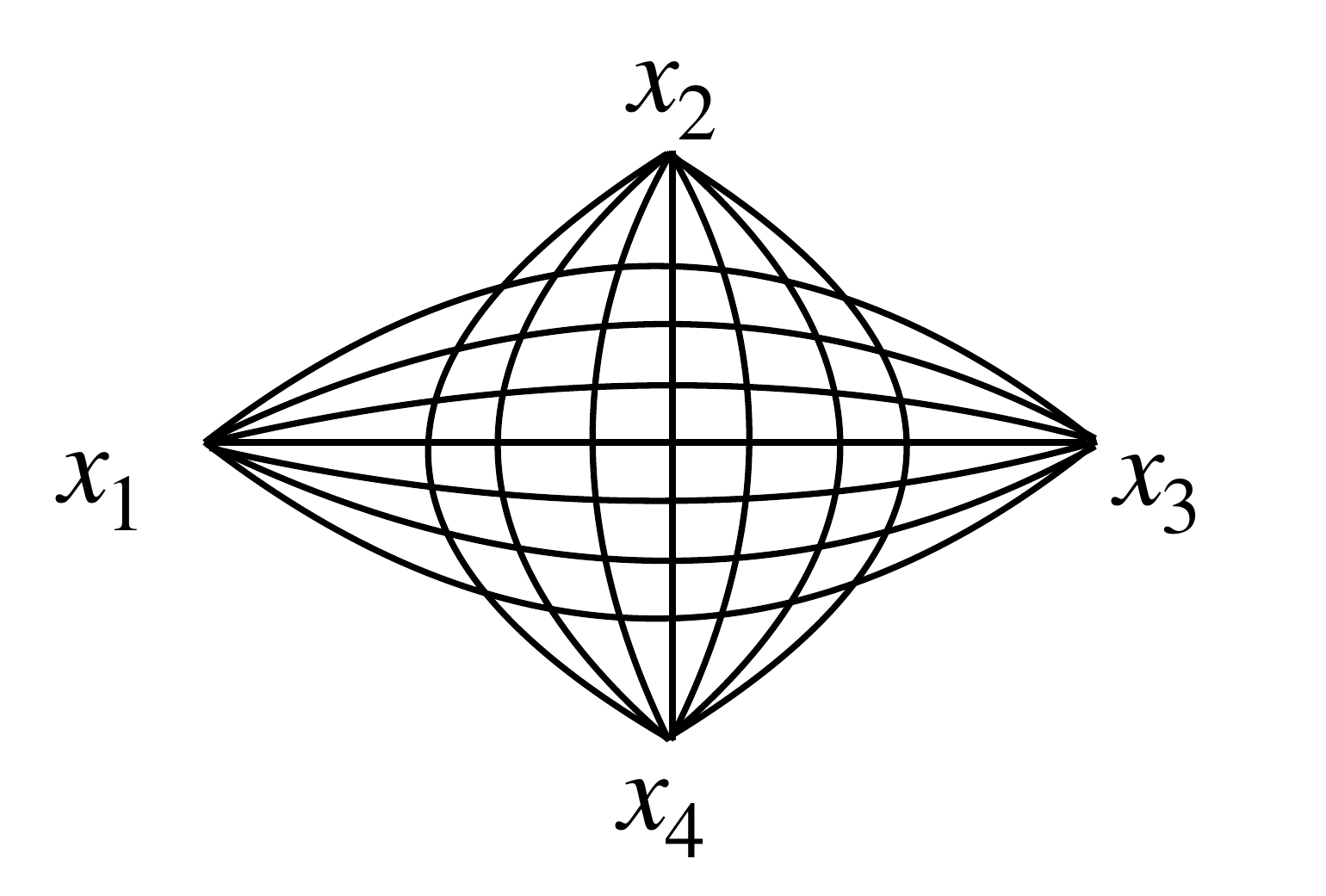}
\end{gathered}
= \frac{g^{2mn}}{(x_{13}^{2})^n (x_{24}^{2})^m} \times
I^{\mathrm{BD}}_{m, n}(z, \bar z)\, , \eee
where $x_{ij}^2 = (x_i-x_j)^2$ and $z, \bar z$ are defined by
  \be
  \begin{aligned}
  \la{crossratios} \U &= {x_{12}^2 x_{34}^2\over x_{13}^2 x_{24}^2} =
  \frac{z \bar{z}}{(1-z)(1-\bar{z})}, \qquad \V= {x_{14}^2
  x_{23}^2\over x_{13}^2 x_{24}^2} ={1\over (1-z)(1-\bar z)}.
    \end{aligned}
  \ee
  A canonical choice for  the positions of the four operators is
 \be \la{canpos} x_1 = (0,0), \ x_2 =(z,\bar z), \ x_3 = (\infty,
 \infty), \ x_4 = (1,1).  \ee
It is convenient to use (for Minkowski kinematics) the exponential
parametrisation
\be \label{param} z = -e^{-\sigma -\varphi }\, , \qquad \bar{z} =
-e^{-\sigma +\varphi  }\, .  \ee
The Euclidean kinematics is described by the analytic continuation of
$\varphi $ to the imaginary axis such that $\bar z = z^*$.  Due to the
symmetries $z\leftrightarrow \bar z$ and $z\leftrightarrow 1/\bar z$
one can consider only the fundamental domain $z\bar z\le 1, \bar z \le
1$, or equivalently $\sigma >0$ and $\varphi >0$.  Basso and Dixon
\cite{Basso:2017jwq} conjectured, using the integrability properties
inherited from ${\mathcal{N}}=4 $ SYM, two different integral
representations for $G_{m,n}$ which they named BMN
(Berenstain-Maldacena-Nastase \cite{Berenstein:2002jq}) and FT
(flux-tube) representations.  The BMN integral representation was
obtained by the adapting the `hexagonalisation'
\cite{BKV1,Fleury:2016ykk,Eden:2016xvg,Fleury:2017eph, Basso:2018cvy}
while the FT representation is obtained by the `pentagon OPE'
\cite{BSV-1,BSV-2}.  A rigorous proof of the former was subsequently
given using the method of separation of variables for conformal
integrable spin chains
\cite{Derkachov:2020zvv,Derkachov:2021rrf,Basso:2019xay,Chicherin:2012yn},
generalising the 2d calculation of \cite{Derkachov:2018rot}.

The BMN integral representation of $G_{m,n}$ looks as the radial part
of an $m\times m$ hermitian matrix integral.  In general, such an
$m$-fold integral can be expressed as an $m\times m$ determinant of
single integrals.  These were identified in \cite{Basso:2021omx} as
the ladder integrals for the case in question.  In particular, the
expression for $G_{1,n}$ is identical to the integral representation
of the ladders found in \cite{Broadhurst:2010ds}.

The thermodynamical limit $m \to \infty$  with fixed cross ratios $\U, \V$ and aspect ratio $n/m$ was was computed in  \cite{Basso:2021omx} using the saddle-point approximation. The open  fishnet, considered as a statistical model
defined on  the $m\times n$ rectangular grid, is charactrerised by an extensive
 free energy   ${\mathcal{F}}=\log  I^{\mathrm{BD}}_{m,n}\sim mn$.
 In \cite{Basso:2021omx} it was observed that the bulk free energy  per site, 
   \be
   \la{defFRBD}
   \hat {\mathcal{F}} =\lim_{m\to\infty} \  {{\mathcal{F}}  \over mn},\qquad 
{\mathcal{F}} \equiv \log  I^{\mathrm{BD}}_{m,n},   \ee
  is different than the   bulk free energy in case of periodic boundary conditions
  computed in \cite{Zamolodchikov:1980mb}.
  Moreover, it is a non-trivial function of the aspect ratio $n/m$. 
   A physical interpretation of this unusual thermodynamical limit due to the boundary conditions is still missing.

A similar phenomenon has been observed in the six-vertex model with
domain-wall boundary conditions \cite{2000JPhA...33.7053K}.  In the
case of the six-vertex model, the dynamics in the continuum limit is
described by a massless bosonic field which in case of domain-wall
boundary conditions has diverging variation on the boundary.  One can
speculate that the rectangular fishnets correspond to singular
boundary conditions for the ADS$_5$ $\sigma$-model.

Since the ADS$_5$ $\sigma$-model has more degrees of freedom than one
massless boson, it is expected that the bulk free energy can be
influenced by the boundary in different ways.  Interestingly, the
authors of \cite{Basso:2021omx} found that in the {\it scaling limit}
combining the thermodynamical limit $m\to\infty$ and the
short-distance limit $x_2\to x_1$, the bulk free energy depends both
on the aspect ratio $n/m$ and on the hyperbolic angle $\sigma =-
\log\sqrt{\U/\V}$.  The relevant scaling parameter is $\hat\s\sim
\s/m$.  The scaling variable $\hat\sigma $ parametrises the flow
between the `bulk' thermodynamical limit ($\hat\sigma \to 0$) and the
Euclidean short-distance limit ($\hat\sigma \to \infty$).  The
saddle-point equation in the scaling limit turns out to be the same as
the finite-gap equation for a classical folded string rotating in
AdS$_3\times S^1$ solved in
\cite{Kazakov:2004nh,Casteill:2007ct,Belitsky:2006en}.  The leading
term in the expansion at $\hat \sigma \to\infty$,
\be \la{stringyFS} \hat {\mathcal{F}}= \log |\hat \sigma | + C_0 +
{C_1\over |\hat \sigma |} + ...  ,\ \ee
describes, in the string theory interpretation, a short folded string
rotating in the flat space, while the subleading terms are corrections
to the flat-space regime coming from the curvature of AdS
\cite{Frolov-Tseytlin-l}.  The analogy with the classical folded
string is intriguing but purely formal.

More generally, one can consider the {\it double scaling limit} $m,
\sigma , \varphi \to\infty$ where the bulk free energy depends on all
three boundary parameters.  It happens that this problem is exactly
solvable and its solution is the main subject of this paper.  The
double scaling limit is characterised by the aspect ratio $n/m$ and
the two scaling parameters
\be \la{doublelight} \hat\sigma \equiv {\sigma \over  2 \pi m}, \
\hat\varphi  \equiv { \varphi  \over  2\pi  m} .\ee
The double scaling limit analytically connects the bulk
thermodynamical limit $(\hat\sigma =\hat\varphi =0$) with the
Euclidean short-distance limit $\hat\sigma \to\infty$ studied in
\cite{Basso:2021omx} as well as with the light-cone limit $\hat\varphi
\to\infty$ to be considered here.  The light-cone limit appears to be
more subtle since the result depends on the ray in the $\{\hat\varphi
, \hat\s\}$ plane along which the infinity is reached.
 
The computation of the free energy is based on the integral
representations obtained in \cite{Basso:2017jwq} and
\cite{Basso:2021omx} which are reminded below to make the presentation
self-consistent.  The so called dual integral, related to the BMN
integral by a Fourier transformation, represents a variant of the
$O(-2)$ matrix model for which resolution techniques have been
established in the last century \cite{Kostov:1992pn}.  For generic
values of the two scaling parameters explicit expression is found not
for the free energy itself but for its derivative with respect to $m$.
This is in principle sufficient to compute to all orders the
corrections to the leading log asymptotics in the short-distance and
light-cone limits, which is however beyond the scope of this short
paper.
   
The paper is organised as follows.  Section \ref{section:finitefish}
presents the derivation of the leading log asymptotics in the
Euclidean short-distance limit ($ \s\to\infty, \varphi =0$) following
\cite{Basso:2021omx}, as well as in the double light-cone limit
($\s=0, \varphi \to\infty$).  The two limits are particular cases of
the single light-cone limit which is characterised by a continuous
parameter $\mu = \s-\varphi $.  Section \ref{sect:doublescalinglimit}
is devoted to the double scaling limit which explore the whole
$\{\hat\s,\hat\varphi \}$ plane.  Since the matrix-model
representation is not symmetric under exchanging $m$ and $n$, it is
more convenient to use as independent variables $m$, which is also the
number of the `eigenvalues', and $\ell \equiv n-m\ge 0$, which enters
the external potential.  The saddle-point equations for a general
potential are reformulated in section \ref{sect:doublescalinglimit} as
a Riemann-Hilbert problem.  In section \ref{sec:Solution} I give the
solution of the R-H problem in terms of incomplete elliptic integrals.
Finally I will show how the leading log asymptotics in the
short-distance and light-like limits are extracted from the general
solution.

 \subsection{BMN integral   representation }

The BMN integral representation, conjectured in \cite{Basso:2017jwq}
and proved in \cite{Derkachov:2019tzo, Derkachov:2020zvv} reads
\be
\begin{aligned}
\la{fishnetintegral} I^{\mathrm{BD}} _{m,n}&=( {\U\V})^{-m/2} \sum_{a_1,...,
a_m=1}^\infty \prod _{j=1}^m {\sinh (a_i \varphi  )\over \sinh\varphi  }\ ( -
1)^{a_j-1} \int \prod _{j=1} ^m {du_j\over 2\pi} e^{2 i \sigma u_j}\
%   {z^{- i u_j  +a_j/2}\bar z ^{- i u_j - a_j/2}
\\
 &\times { \prod_{i=1}^m a_i \prod_{i<j} \left[ (u_i - u_j)^2 +
 {(a_i+a_j)^2 \over 4}\right] \left[ (u_i - u_j)^2 + {(a_i-a_j)^2
 \over 4}\right]
%\Delta_{a_i,a_j}(u _i, u_j)
\over  \left(u _j^2 +{ a_j^2/4}\right)^{m+n}} 
 ,
\eee
where $({\U \V })^{-1/2} = 2 \cosh \sigma +2 \cosh \varphi  $ in the
exponential parametrisation \re{param}.  The Euclidean kinematics is
attained by the analytic continuation of $\varphi  $ to the imaginary axis
such that $\bar z = z^*$.

The BMN integral is very similar to the expansion of the octagon
\cite{Coronado:2018ypq} in a series of multiple integrals at weak
coupling.  More precisely, the BMN integral is obtained by retaining
the term with $m$ virtual particles and taking the weak coupling limit
of the weights.  The representation of the octagon with bridge $\ell=
n-m$ in terms of free fermions \cite{Kostov:2021omc} implies a similar
representation for the fishnet, which is spelt out in appendix
\ref{app:FF}.

\subsection{ The Fourier transformed integral}

In \cite{Basso:2021omx}, the BMN integral \re{fishnetintegral} was
given, using the determinant representation in terms of ladders
obtained in \cite{Basso:2017jwq}, a dual form
 \bee \la{MMdef} I^{\mathrm{BD}} _{m,n}& = {\mathcal{Z}}_{m}(\ell ,
 \sigma , \varphi ) , \qquad \ell \equiv n-m.  \eee
 Up to a normalisation factor, the dual integral takes the form of the
 partition function of the $O(-2)$ matrix model \cite{Kostov:1992pn}
\bee \la{MMpartf} {\mathcal{Z}}_{m}(\ell , \sigma , \varphi  ) & = {1\over
\mathcal{N} } \, {1\over m!} \int_{|\sigma |}^\infty \prod_{j=1}^m \
dt_j \ e^{-V(t_j)} \ \prod_{j,k =1}^m (t_j+t_k) \prod_{j<k}^m
(t_j-t_k)^2 \eee
with particular interaction potential
\bee \la{potentialgen} V(t) &= \log { \cosh t + \cosh \varphi \over
\cosh \sigma + \cosh \varphi } - \ell \log(t^2-\sigma ^2) .
\la{dpoten} \eee
  In the matrix-model interpretation, the integration variables
  $t_1,..., t_m$ the eigenvalues of a Hermitian $m\times m$ matrix.
  The normalisation factor reads
   \be \la{defNNN} \mathcal{N} = \prod_{i=0}^{m-1} ( 2 i+\ell )!  (2
   i+1+\ell)!
     =  {G(2m+\ell +1)\over G(\ell+1)},
      \ee
where $G(m)\equiv {\tt BarnesG}[m] = 1!2!...  (m-2)!  $ is Barnes'
G-function.  In appendix \ref{app:FF} I show that the dual integral is
obtained from the BMN integral \re{fishnetintegral} by a Fourier
transformation.  The spectral variable $t$ is therefore the `momentum'
conjugated to the rapidity $u$.

\subsection{Determinant of ladders }

The rhs of \re{MMpartf} gves, for $m=1$, the integral representation
of the ladder integrals obtained earlier by Broadhurst and Davydychev
\cite{Broadhurst:2010ds},
 \be \la{deffkin} \begin{aligned} f_k(z, \bar z) &\equiv {(1-z)(1-\bar
 z)\over z-\bar z} k!(k-1)!  L_k(z, \bar z) = \int_{|\sigma |
 }^{\infty } { \cosh \sigma +\cosh \varphi \over \cosh t +\cosh
 \varphi } \, (t^{2}-\sigma ^2)^{k-1} \ 2 t dt \\
& = (1-z)(1-\bar z) \sum_{j=k}^{2k} \frac{(k-1)!  \ j!  }{ (j-k)!
(2k-j)!} \ (-\log z\bar z)^{2k-j} \ {\mathrm{Li}_{j}(z) -
\mathrm{Li}_{j}(\bar z)\over z-\bar z} .
\end{aligned}
\ee
 The general $m$, the integral \re{MMpartf} is equivalent to the
 original Basso-Dixon determinant representation of the fishnet
 \cite{Basso:2017jwq}.  The latter is obtained by writing the product
 in the integrand in \re{MMpartf} as \bee \prod_{i=1}^m (t_j^2-\sigma
 ^2)^\ell \prod_{j<k}^m \left( (t_j^2 -\sigma ^2)-( t_k^2 -\sigma
 ^2)\right) ^2 &=(t_j^2-\sigma ^2)^\ell \left( \det_{j,k=1}^m \left[
 (t_j^2 -\sigma ^2)^{k-1 }\right]\right)^2 \\
		&= \det_{j,k=1}^m \left[ \sum_{i=1}^m (t_i^2-\sigma
		^2)^{j+k-2+\ell} \right] \eee
 which leads, via the Cauchy-Binet formula, to
\be \la{detrep} I^{\mathrm{BD}} _{m,m+\ell} = { 1\over {\mathcal{N}}}
\det\left( \left[ f_{j+k+\ell-1}\right]_{j,k=1,..., m}\right) \ .  \ee

\section{Short-distance and  light-cone  limits for finite  fishnets}
\la{section:finitefish}

\subsection{Euclidean short-distance limit \ ($\sigma \to \infty$ with
$\varphi  $ fixed)
%($z\to 0, \bar z\to 0$) $(\sigma \to\infty,\ \mu= |\varphi  |- |\sigma |
%\to -\infty$)
}

If $ \sigma \to\infty$, then
  %$(z, \bar z) \to (0,0) $,
$\U\to 0$ and $\V\to 1$.  As ${x_{14}^2 x_{23}^2 = x_{13}^2
x_{24}^2}$, the limit $\U \to 0$ implies that either $x_3\to x_4$ or
$x_1\to x_2$.  By conformal transformation one can achieve that both
conditions are satisfied, with
  \bee \la{x1234zO} |x_{12}|^2 , |x_{34}|^2 \sim \sqrt{ { \U}} \
  |x_{13}|^2 \, \qquad ( \U\to 0, \ \V\to 1), \eee
hence $x_1\sim x_2$ and $x_3\sim x_4$.  This is the Euclidean
short-distance, or OPE, limit studied in \cite{Basso:2021omx}.  For
the sake of completeness, I sketch the derivation of the leading log
asymptotics given there.
 
 The short-distance limit is achieved by sending $\sigma \to\infty$
 with $\varphi  $ finite.  The ladder integrals \re{deffkin} become (after
 shifting the integration variable $t\to t-\sigma $)
\be \la{fEOPE} f_k(z, \bar z) \ \underset{\sigma \gg k} \to\
   \int_{0} ^\infty (2|\sigma |)^{k }\ t^{k-1 } e^{-t} dt&= (2|\sigma
   |)^{k } \ (k-1)!  \ee
and the determinant formula \re{detrep} gives \cite{Basso:2021omx}
 \bee \la{MMdefMOPE} I^{\mathrm{BD}} _{m,m+\ell} & \to\ { (2|\sigma
 |)^{m (m+\ell)} \over {\mathcal{N}}} \det _{j,k}\left[ (j+k+\ell
 -2)!\right] = \left(\log {1\over \U}\right)^{m (m+\ell)} \
 C_{m,m+\ell} \eee
where 
\be \la{defCmn} C_{m,n} = {G(m+1)G(n+1)\over G(m +n+1)}, \quad G(m) =
1!2!...  (m-2)!\ .  \ee

\subsection{Double light-cone, or null, limit ($\varphi \to\infty$
with $\sigma $ fixed)}

Sending $\varphi  \to\infty$ with $\s$ finite implies $\{\U,\V\}\to \{ 0,
0\}$, which means that the Minkowski intervals (but not the Euclidean
distances!)  $x_{12}^2$ and $x_{14} ^2$ become simultaneously near
light-like,
 \bee \la{x1234z} x_{12}^2 , x_{34}^2 \sim \sqrt{{ \U}} \ |x_{13}||
 x_{24}| , \quad x_{14}^2 , x_{23}^2 \sim \sqrt{{ \V}} \ |x_{13}|
 |x_{24}| \, .
  \eee
 The ladder integrals $f_k$ can be approximated by
 \be f _{k } (z,\bar z) \underset{ \varphi  \gg k} \to\ 2 \int_0^\varphi 
 t^{2k-1} dt = \frac{\varphi  ^{2 k}}{k}
 \ee
 and the determinant formula gives, for $m, \ell \ll \varphi  $,
  \bee \la{MMdefMOPEll} I^{\mathrm{BD}} _{m,m+\ell}&
  \to { \varphi  ^{2m(m+\ell)} \over \mathcal{N} } \det \left[ {1\over
  i+j-1+\ell}\right]_{i,j=1,..., m}
.
  \eee
  To compute the determinant, denote $x_i= i , y_j= j+\ell-1$ and
  apply Cauchy's identity \footnote{I thank Philippe Di Francesco for
  suggesting that.},
 \bee & \det \left[ {1\over i+j-1+\ell}\right]_{i,j=1,..., m} =
 \det_{i,j=1,...,m} {1\over x_i+y_j}={ \Delta(x) \Delta(y)\over
 \prod_{i,j=1}^m (x_i+y_j) } \\
&= \frac {\prod_{i=1}^{m-1} i!^2}{\prod_{i,j=1}^m (i+j+\ell-1)}=
{G(m+1) ^2 G(m+\ell+1)^2\over G(2m+\ell+1) G(\ell+1)} = {\mathcal{N}}\, (
C_{m,m+\ell})^2.  \eee
Since $\varphi  ^2 = \log \U \log\V$, the fishnet integral takes in the
double light-cone limit the following factorised form,
\bee \la{DLLscaling} I^{\mathrm{BD}} _{m,n+\ell} &= C_{m,m+\ell}\,
\left(\log{1\over \U} \right)^{m(m+\ell)} \times C_{m,m+\ell}\, \left(
\log {1\over \V}\right)^{m(m+\ell)} \, .  \qquad \eee
The two factors are obviously associated with the two pairs of
operators which become light-like.  A factorised expression very
similar to \re{DLLscaling} was recently obtained in
\cite{Arkani-Hamed:2022cqe} for the leading log singularities of the
dimensionally regularised Basso-Dixon fishnet in momentum space.

\subsection{Single light-cone limit ($ \varphi \to\infty $ with
$\mu\equiv \varphi - \sigma $ fixed)}

The limit $\varphi \to\infty$ with $\mu= \varphi -\s$ fixed, or $z\to 0$ with
fixed $\bar z=e^{\mu}$, translates in terms of $\U$ and $\V$ as
  \be \la{LCL-UV } \U\to 0, \ \ \V\to {1\over 1+ e^\mu} \ \qquad
  (\mu\equiv \varphi  -\sigma ).  \ee
   For $\mu$ finite, the sides 12 and 34 are close to light-like, 
  \bee \la{x12-34z} x_{12}^2 , x_{34}^2 \sim \sqrt{{ \U}} \ |x_{13}||
  x_{24}| , \eee
  while
the sides 23 and 42 remain in general position.

It is obvious from \re{LCL-UV } that the light-like limit interpolates
continuously between the Euclidean short-distance limit $(\mu \to
-\infty)$ and the double light-like limit $(\mu\to +\infty)$.  The
dependence on $z$ and $\bar z$ of the ladder integral \re{deffkin}
factorises,
\bee \la{fkLC} f_k(z,\bar z) & \to \left( 2|\sigma |\right)^k \int_0^\infty
{1+e^\mu \over e^t+e^\mu}\ t^{k-1} dt \\
& = -(k-1)!  \, \left(\log{1\over z} \right)^k \left(1 -{1\over \bar
z}\right)\, { \mathrm{Li}}_k( \bar z)
\eee
hence the fishnet integral factirises as well.  The dependence on $z$
exhibits the standard leading log singularity, while the dependence on
$\bar z$ is more involved,
  \bee \la{MMdefMOPEsl} I^{\mathrm{BD}} _{m,m+\ell}&\underset{{ z\to 0}}\to {
  \left(\log {1\over z }\right) ^{m(m+\ell)} } \
\
F (\bar z) \, ,
%\\ F_{m,m+\ell} (\bar z)&={1\over \mathcal{N} } \left({ \bar z-1\over
%\bar z}\right)^m { \det _{i,j}^m\left[  (i+j-2+\ell)!  { \mathrm{Li}}_{i+j-1+\ell}(\bar z)
%\right] \over \det _{i,j}^m\left[  (i+j-2+\ell)!  \right]}.
  \eee
with $F(\bar z)$ being an $m\times m$ determinant of polylogs.

 \section{Double scaling  limit}
 \la{sect:doublescalinglimit}

 Now let us consider the most general double scaling limit achieved by
 combining thermodynamical limit $m\to\infty$ with the scaling
 \re{doublelight} of the cross ratios $u $ and $ {v}$,
 \be \la{doublelightb} \ell, m, \sigma , \varphi  \to\infty\ \ \text{with}\
 \ \hat\sigma \equiv {\sigma \over 2\pi  m}, \ \hat\varphi  \equiv {
 \varphi  \over  2\pi  m} ,\ \hat \ell \equiv {\ell\over  2\pi  m}\
 \ \text{ fixed}.  \ee
 The third parameter $\hat \ell$ controls the aspect ratio, $n/m = (1+
 \hat \ell)/(1-\hat \ell)$.  The so defined scaling parameters
 $\hat\varphi $ and $\hat\s$ are invariant under exchanging
 $m\leftrightarrow n$ while $\hat \ell $ changes sign.  I will stick
 most of the time to the original non-normalised variables $\ell,
 \sigma , \varphi $, assuming the scaling \re{doublelightb}, i.e.
 $\ell, \sigma ,\varphi \sim m$ with $m$ sufficiently large.  This
 will make more obvious the comparison of the results obtained for
 different scales.

The goal is to compute the leading contribution to the free energy
$\mathcal{F} _m=\log{\mathcal{Z}}_m$, or more strictly the scaling
function which defines free energy per vertex
 \be \la{defFE} { \mathrm{Li}}m _{m\to\infty} { {\mathcal{F}}
 _m(\ell,\sigma ,\varphi )\over m (m+\ell) } = \, \hat {\mathcal{F}}
 (\hat \sigma , \hat\varphi , \hat\ell).  \ee

\bigskip

The multiple integral \re{MMpartf} describes a statistical ensemble of
$m$ identical particles characterised by a repulsive two-body Coulomb
interaction and a confining potential \re{potentialgen},
 \be \la{Potentialsplit} V(t) =V_0(t)- \ell \log(t^2-\sigma ^2) ,
 \quad V_0(t)= \log { \cosh t + \cosh \varphi  \over \cosh \sigma + \cosh
 \varphi  } .  \ee
In the thermodynamical limit, the fluctuations are suppressed and the
partition function is determined by the configuration minimising the
energy.  The positions $t_1>...>t_m\ge |\sigma |$ of the $m$ particles
at equilibrium are determined by the saddle-point equations
\bee
\la{sdpteqs}
{\partial {\mathcal{S}} / \partial  t_i}& =0  \qquad (i=1,..., m )
\eee
where ${\mathcal{S}}$ is the  total energy  
 \bee \la{defCS} {\mathcal{S}}&\equiv \sum_{j=1}^mV(t_j) - \sum_{k\ne j} ^m
 \log(t_k^2-t_j^2) -\sum_{j=1}^m \log (2 t_j) .  \eee
 The saddle-point equations read explicitly
 \bee \la{sdpteqsA} V'(t_j)& = \sum_{k\ne j} ^m {2\over t_j-t_k} +
 \sum_{k=1}^m {2\over t_j+t_k} =0 \qquad (i=1,..., m ).  \eee

With the potential \re{MMpartf}, the roots $t_j$ of the saddle-point
equations are real and scale as $t_j\sim m$.  In the thermodynamical
limit the sum can be approximated by an integral with a continuous
density $\rho(t)$.  At macroscopic scale the density is supported by a
compact interval $[a,b]$ with $|\s| \le b<a <|\varphi |$.  It is
useful to extend the density to the whole real axis by the symmetry
$\rho(t)= \rho(-t)$ and consider the saddle-point solution as a
symmetric distribution of $2m$ particles with support $
[-a,-b]\cup[b,a] $.  Then the saddle-point equations take the form (at
macroscopic scale)
 \be \la{sdpeqdens} V'(t) = 2 -\hskip -0.4cm \int_{\mathbb{R}} {d t'
 \rho(t')\over t-t'} \qquad ( b<|t|<a).  \ee
Obviously $a,b\sim m$ while $\rho$ remains finite when $m\to\infty$.

The standard technique to solve the saddle-point equations is by
reformulating the integral equation \re{sdpeqdens} as a
Riemann-Hilbert like problem.  For that introduce the resolvent
 \bee G(t)&= \sum_{k=1}^m{1\over t- t_k} = \int _ b^a {dt'
 \rho(t')\over t-t'} \eee
and the function (giving the force acting on a probe particle at the
point $t\in\IC$) \bee \H (t)&
%          \equiv -\hf 
%          \partial _t\partial hi (t) 
= - \hf V'(t) + G(t)- G(-t).  \eee
The meromorphic function $H(t)$ has large-$t$ asymptotics
 \be \la{asympJ} \H (t) = - \hf V'(t)+ {2m \over t}+ O(t^{-3}), \ee
and, apart from the singularities inherited from the external
potential, two cuts $[-a, -b]$ and $[b,a]$ on the real axis.  The
integral equation \re{sdpeqdens} can be formulated as the boundary
condition
\be \la{bcH} {H(t-i0)-H(t+i0)\over 2\pi i} \times \big(
H(t+i0)+H(t-i0)\big) =0, \qquad t\in {\mathbb{R}}, \ee
As a consequence, the square $H^2(t)$ is analytic in the vicinity of
the real axis which, together with the asymptotics \re{asympJ} at
$t\to\infty$ determines the function $H(t)$ uniquely.  By a standard
argument\footnote{For the rescaled variable $\hat t=t/2\pi m$ the
width of the analyticity strip vanishes as $1/m$ because of the arrays
of poles at $\Re t = \pm \varphi $ of $V'(t)$, but these poles are in
general at macroscopic distance from the branch points and the
standard argument still works.  } (see e.g. section 2.1 of
\cite{DiFrancesco:1993nw}), the function $\H (t) $ can be written as a
linear integral
 \bee 
 \la{origint} \H (t) &= \dashint _b^a {dt_1 \over 2\pi}{ 2t\,
 V'(t _1 ) \over t^2-t_1 ^2} {y(t)\over y(t_1)} =-2 \int _b^a {dt_1
 \over 2\pi}\ {y(t)\over y(t_1)}\ { t V'(t) -t_1\, V'(t_1 )\over
 t^2-t_1 ^2} \ , \eee
where $y(t)$ is the positive root of the equation
\bee \la{defRES} a^2 y^2= (a^2- t^2)(t^2 - b^2) .  \eee
The endpoints $a$ and $b$ of the eigenvalue distribution are
determined by the asymptotics at infinity \re{asympJ} which imposes
the constraints
\bee \la{asymcds} & \int _b^a {dt \over y(t)} V'(t) =0, \quad \int
_b^a {dt \over y(t)} t^2 V'(t) = { 2 \pi m} a.  \eee

For the computation of the free energy it is useful to introduce the
{\it effective potential } $\phi(t)$ of a probe particle at the point
$t\in \IC$,
\bee \phi(t) = \phi(-t) &= V(t) + 2\int_t^\infty G(t) dt = 2
\int_t^\infty H(t) dt.  \eee
Obviously the effective potential must be constant on the two cuts,
\be \la{defphio} \phi(t) = \phi_0, \quad b\le |t|\le a.  \ee
The constant $\phi_0 = \phi(a)$ is an important collective variable.
It gives the increase of the critical action upon bringing a new
particle from infinity and therefore is conjugate to the number of
particles $m$,
\be \la{phidFm} \partial_m {\mathcal{S}}_c= \phi_0.  \ee

The constant $\phi_0$ can be computed once the expression for the
derivative $\partial_m H(t)$ is known.  By \re{asympJ}, the derivative
$\partial_m H(t)$ behaves at infinity as $2/t$ and thus depends on the
external potential only through the positions of the branch points.
It defines a normalised Abelian differential of first kind on the
elliptic curve with equation \re{defRES} with singular point at
infinity and has a standard form
\bee \la{defomega} d \omega(t) &= \partial_m \H (t)\, dt = \frac{ 2t
dt}{\sqrt{t^2-a^2} \sqrt{t^2-b^2}} , \quad \omega(t) =
2 \log {\sqrt{t^2-a^2}+\sqrt{t^2-b^2}\over 2} .  \eee
 Now let us compute how saddle-point energy ${\mathcal{S}}_c$ changes
 with the number of particles.  Starting with \re{defCS}, the
 saddle-point energy can be written as an integral with the density,
\begin{align}
{\mathcal{S}} _c &= \int _{b}^a dt \rho(t) \Big(V (t) \, + \hf[ \phi(t)-
V(t)]\Big) = \hf \int_b^a dt \, \rho(t)\, V(t) + \hf m\, \phi_0 ,
\la{fren2int}
\end{align}
where the equilibrium condition \re{defphio} have been used.
Combining the derivative of \re{fren2int} in $m$ and \re{phidFm},
$\phi_0=\phi(a)$ can be expressed in terms of the Abelian differential
\re{defomega} as
\bee \la{difeqpmF} \phi_0 & = m \partial_m \phi(a) +\int_b^a dt
\partial_m\rho(t) V(t) \\
 &= - 2 m \, \omega(a) +{1 \over \pi a} \int\limits_{ b} ^a {2t dt
 \over y(t)} V(t) .  \eee
From here one obtains for the first derivative of the free energy
\bee \la{dmF} \partial_m\mathcal{F} &=-\partial_m \log {\mathcal{N}}
-\phi_0 \\
 &=- \partial_m \log{\mathcal{N}} + 2 m \log {a^2-b^2\over 4} -{
 1\over\pi}\int_b^a dt \frac{ 2t \, V(t)}{\sqrt{a^2-t^2}
 \sqrt{t^2-b^2}} \, .  \eee
The integral in the last term is relatively easy, unlike the integral
in \re{fren2int}.  Differentiating once again, one obtains for the
second derivative of the free energy
\bee\la{dphia} \partial_m^2\mathcal{F} &= - \partial_m^2 \log
{\mathcal{N}}-\partial_m\phi (a) = 2 \log{a^2- b^2\over 4 (2 m+\ell)
^2}
   .
 \eee
For the free energy itself,
\bee \la{Fdsl} \mathcal{F} 
%      &= -\hf m\log (\U \V ) -\log {\mathcal{N}} - {\mathcal{S}}_0 \\
&= -\log {\mathcal{N}} - \hf \int_b^a dt \rho(t)\, V(t) - \hf \phi_0 , \eee
explicit formulas can be obtained only at particular points in the
parameter space characterised by single scaling limits.

 \section{Solution of the saddle-point equations in the double scaling
 limit} \la{sec:Solution}

At macroscopic scale $t\sim m$, the piece $V_0$ in \re{Potentialsplit}
as a function of the complex variable $t$ is approximated by
	 \bee \la{LeadingPot} V_0(t) =
	  \max\left( |t|, |\varphi|\right) -  \max\left( |\sigma|, |\varphi |\right)
, 
  \ \  V'(t) =\mathrm{sgn} (t)\, \theta\left(|t|-|\varphi   |\right) 
 \qquad   (t\in{\mathbb{R}})
 \eee
 where $\theta(t)$ is the Heaviside function.  The extension to the
 complex plane is done by simply replacing $t\to \Re t$ in
 \re{LeadingPot}.  Since at macroscopic scale the derivative of the
 external potential develops a discontinuity $t= \pm \varphi $, the
 solution for the saddle-point equations have different analytical
 properties depending on whether the point $\varphi $ belongs to the
 support of the equilibrium density.  The parameter space splits into
 two domains depicted in fig.  \ref{fig:phasediagram},
 
 I. \ $b\le  |\varphi | <a$
 
  II. $  |\varphi |  \le b <a$.
 
\noindent In the domain I the density is a smooth function while in
the domain II it develops a logarithmic cusp.
       
   \bigskip   
  
 \noindent $ \bullet$ \ {\it Regime I}\ \ ($ |\varphi  | <b$)
  
  \smallskip
  
 \noindent If $|\varphi |<b$, the saddle-point equations do not depend on
 the parameter $\varphi  $.  In \cite{Basso:2021omx} it was noticed that in
 this regime the integral equation \re{sdpeqdens} coincide with the
 finite-gap equation
 \cite{Kazakov:2004nh,Casteill:2007ct,Belitsky:2006en} for the
 Frolov-Tseytlin folded string \cite{Frolov-Tseytlin-l} rotating in
 AdS$_3\times S^1$ with quantum numbers $\{S,J\} =\{2m,\ell\}$ and
 $\s\sim $ string tension.  The saddle-point equations \re{sdpteqsA}
 can be obtained starting with the Bethe equations
 \be \la{beteeqs} \left( {t_j - {\sigma^2\over t_j}+ i\pi \over t_j -
 {\sigma^2\over t_j} - i\pi }\right)^\ell \prod _{k\ne j}^{2m} {t_j-t_k + 2\pi
 i \over t_j-t_k - 2\pi i } =1, \qquad (j=1,..., 2m) \ee
by assuming that all roots are large and imposing the constraint that
the configuration of the roots is even, $ \{t_j\} = \{- t_j\}$.  The
solution is completely characterised by the choice of the mode numbers
which define the branch of the logarithms when the Bethe equations are
written in logarithmic form.  With the symmetry of the roots taken
into account, the equations for the positive roots read
\be \la{BAElog} {2\ell\, t_j\over t_j^2 - \sigma^2}+ \sum_{k\ne j} ^m
{2\over t_j-t_k} + \sum_{k=1}^m {2\over t_j+t_k} = n_j\qquad (j=1,...,
m).  \ee
The folded string solution corresponds to choosing $n_1=...=n_m=1$ and
taking the limit of large charges $m$ and $\ell$.

As the free energy is sensitive to the constant mode of the piece
$V_0$ of the potential, eq.  \re{LeadingPot}, the regime I splits into
two subdomains separated by the lines $|\s|=|\varphi |$,

\bigskip
 
 -- Regime Ia\ \ ($ |\sigma |\le |\varphi  | \le b <a$)
 
 -- Regime Ib\ \ ($ |\varphi  | \le |\sigma | \le b <a$).

   \bigskip   
  
 \noindent   $ \bullet$  {\it Regime  II}\ \ ($  b<|\varphi   |  <a$) 
  
  \smallskip
  
\noindent If $|\varphi |>b$, the non-analyticity of the external potential
at $t=\varphi $ affects the saddle-point equations and the spectral density
develops a cusp at $t=\varphi  $.  To get more intuition about the origin
of the cusp, let us turn to the interpretation of the saddle-point
equations as Bethe equations for the generalised $sl(2)$ spin chain,
eqs.  \re{beteeqs}-\re{BAElog}.  Regime II is again characterised by
an even distribution of the Bethe roots, but with different choice for
the mode numbers, namely $n_ j=1$ if $ t_j>|\varphi |$ and $n_j=0$ if $
t_j<|\varphi |$.  This choice for the mode numbers is formally allowed but
does not describe a finite gap solution in the limit of large charges.
The two groups of eigenvalues characterised by mode numbers $1$ and
$0$ do not condense to separated cuts but instead collide at the point
$t=\varphi $ producing a logarithmic cusp of the density.
 
Although the solutions of the saddle-point equations in regime II and
in regime I are not related analytically, they match continuously on
the curve with equation $b=|\varphi |$ which separates the two regimes.
Since the solution in the regime I does not not depend on $\varphi $, it
can be computed for $\varphi =b$.  In this sense the regime I is contained
in the regime II as a boundary value.

   \bigskip

\begin{figure}[tbp]
\centering 
\includegraphics[width=0.56\textwidth]{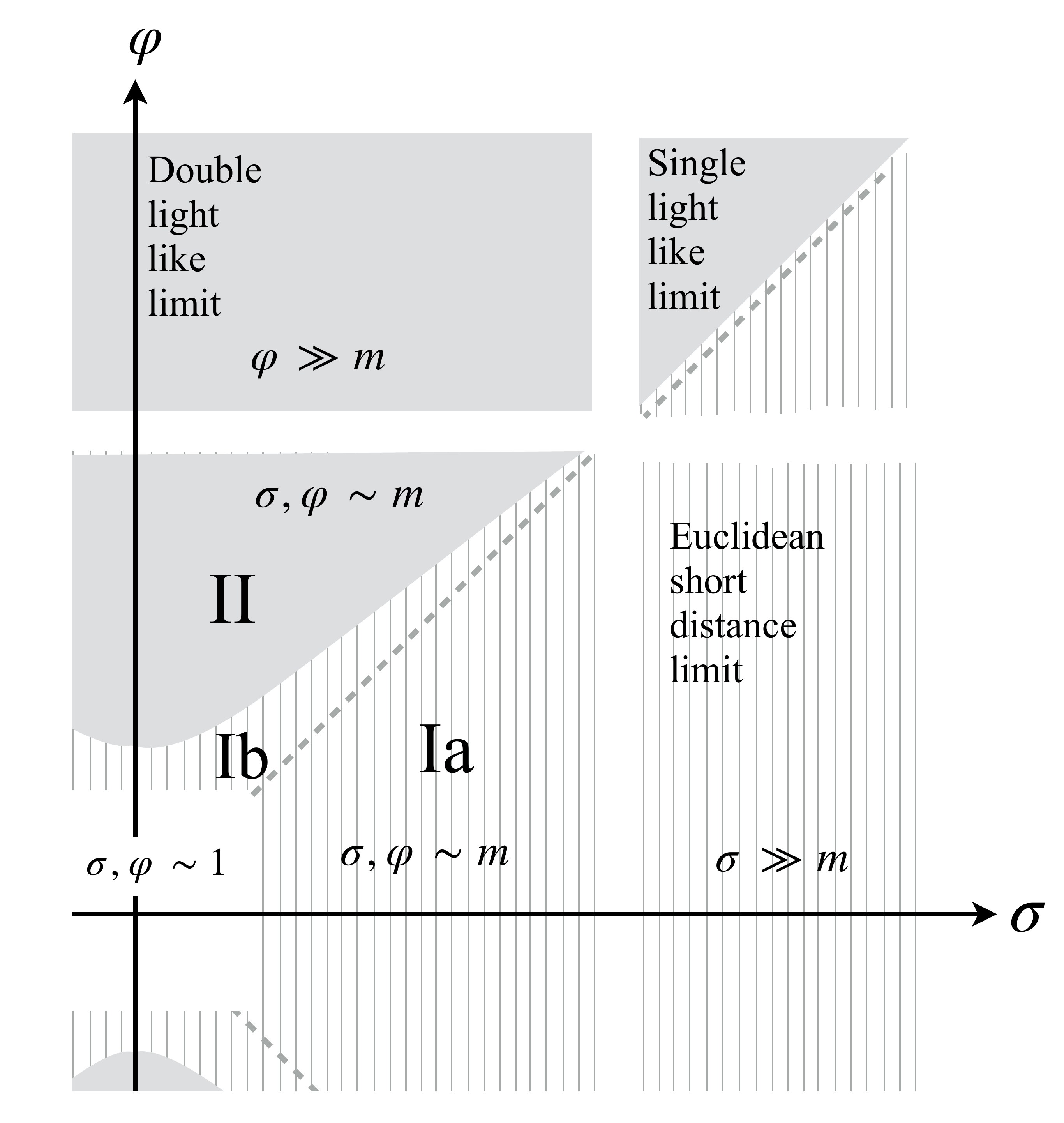}

\caption{\small The scaling regimes of a large rectangular fishnet.
The double scaling domain $\s,\varphi \sim m$ is split into regimes I
and II. In regime I (the hatched area $|\varphi |\le b$)) the solution
for the spectral density does not depend on $\varphi $.  In the domain
$|\varphi | <|\s|$ (regime Ia) this is so also for the free energy
while for $|\s|< \varphi \le b$ (regime Ib) the free energy contains a
term $m\varphi $.  The solution in regime I relates analytically the
bulk thermodynamical limit c $\s,\varphi \sim 1$ and the Euclidean
short distance limit $\sigma \gg m$.  In regime II (the grey domain
$|\varphi |\ge b$) the solution depends on both $\sigma $ and $\varphi
$.  The double light-cone limits is attained in regime II by taking
$\varphi \gg m$ with $\s$ finite.  The single light-cone limit is
reached by taking $\varphi \gg m$ with $\mu =\varphi -\sigma $ fixed.
}
   
\label{fig:phasediagram}
\end{figure}

\subsection{   Solution in regime I}

The solution in regime I has been found in
\cite{Kazakov:2004nh,Casteill:2007ct,Belitsky:2006en} and adjusted for
the fishnet in \cite{Basso:2021omx}.  The equations for the branch
points, eq.  \re{asymcds}, are expressed in terms of the complete
elliptic integrals of first and second kind $\mathbb{E}=E(k^2)$ and $
\K=K(k^2)$ (the notations for the elliptic integrals are collected in
appendix \ref{app:Elliptic}), namely
\bee \la{dmFFBDff1} \sqrt{ (a^2- \sigma ^2 )( b^2 - \sigma ^2 )} \,
{\K} &= { \pi \ell\, a } ,\\
 {a^2 \mathbb{E} } -\sigma^2 \mathbb{K} &= \pi(2 m+\ell )a ; \\
  k^2&=1- k'^2,   \ \ k' = {b /a}.
 \eee
The spectral density is expressed in terms of the complete elliptic
integral of third kind,
\bee \rho(t) = & {1 \over\pi} { \ell \, t\over t^2-\sigma ^2}
\sqrt{{(a^2-t^2)(t^2-b^2)\over (a^2-\sigma ^2)(b^2-\sigma ^2)}} + {1
\over \pi^2} {t\over a}\sqrt{t^2-b^2\over a^2-t^2}\ \Pi\left(
{a^2-b^2\over a^2-t^2}\Big| 1- {b^2\over a^2}\right).  \eee
 For the derivative of the free energy one finds from \re{dmF} 
\bee \la{dFFFSHDgen} \partial_m \mathcal{F}  &= (2 m+\ell) \log{(a^2-b^2 )\over 4
(2m+\ell)^2 } + 2 \ell \, \arctanh{\sqrt{b^2-\sigma
^2}\over\sqrt{a^2-\sigma ^2}} \\
 &- \frac{ 2\ell \, \sigma ^2 }{\sqrt{ \left(a^2-\sigma ^2\right)
 \left(b^2-\sigma ^2\right) }} +\max( |\varphi | , |\sigma ), \eee
 where   the asymptotics 
\bee \partial_m \log{\mathcal{N}}&\approx 2 (2 m+\ell ) \log {2 m+\ell
\over e} \la{asymNN} \eee
is taken into account.  There is no reasons to believe that the
integration of \re{dFFFSHDgen} can be done explicitly for general
$\s$.  The special cases $\s=0$ and $\s\to\infty$ have been solved
completely in \cite{Basso:2021omx}, to be reviewed below.

 \medskip
 
\noindent $\bullet$ {\it Bulk thermodynamical limit ($\sigma=0 $)}
   
This case correspond to the origin in fig.  \ref{fig:phasediagram} The
equations for the branch points are
 \bee {\E\over\pi} &= {2m+\ell\over a} , \quad {\K\over \pi} = {
 \ell\over b} \, \qquad (\sigma =\varphi  =0) \eee
and the resolvent is essentially Heuman's Lambda function $
\Lambda_0(\psi , k^2) $,
\bee \la{ResolventBulk} H(t)&= \hf \Lambda_0\left(\arcsin{a\over t},
k^2\right) \, - \hf \mathrm{sgn} (\Re t), \\
\rho(t) & = - {1\over 2\pi} \Im\Lambda_0\left(\arcsin{a\over t}, k^2\right).
\eee
The derivative of the free energy,
\bee \la{dFFFSHDgena} \partial_m \mathcal{F} &= (2 m+\ell)
\log{(a^2-b^2 )\over 4 (2m+\ell)^2 } + 2 \ell \, \arctanh ({b / a }),
\eee
 integrates to   \cite{Basso:2021omx} 
\bee \la{FRENBB} \mathcal{F} & = m^2 \log \frac{a-b}{2} +(m+\ell )^2
\log \frac{a+b}{2}- {\ell ^2 \over 2}\log {a b\over \ell} - {(2 m+\ell
)^2 \over 2}\log (2 m+\ell ) \\
  & =
   m^2 \log {1-k'\over 2}
+n^2 \log   {1+k'\over 2}+2 m n \log  \pi 
   \\
&+ \hf (m-n)^2 \log \K-\hf (m+n)^2 \log \mathbb{E} .
 \\
\eee
The constant of integration was determined by the requirement that the
free energy must vanish when $m=0$.

\medskip \noindent $\bullet$ {\it Euclidean short-distance limit
$\sigma \gg m$}

When $\sigma \gg m$, the positions of the branch points scale as
$a-\s\sim m, \b-\s\sim m$ so that $k^2\sim 1/\s$ and the elliptic
curve \re{defRES} degenerates into a gaussian one.  Setting
 \bee a &= \sigma + S+2R, \ b = \sigma + S-2R , \eee
where $S,R\sim m$, and expanding \re{dmFFBDff1} at large $\sigma $,
one obtains in the leading order
  \bee R&= \sqrt{m (m+\ell )} ,\ S = 2m+\ell \qquad k^2 =1-b^2/a^2
  \approx { 8R/\sigma } .  \eee
 One obtains for the leading large $\sigma $ asymptotics the free
 energy
\bee \la{FFSDA} \mathcal{F} &= m (m+\ell ) \log (2 \sigma
)+\frac{3}{2} m (m+\ell ) +\hf m^2 \log (m) \\
&+\hf (m+\ell )^2 \log (m+\ell ) - \hf (2 m+\ell )^2 \log (2 m+\ell )
\qquad(\s\gg m,\ell\gg 1).
  \eee
The expression \re{FFSDA} matches the large $m$ asymptotics of
\re{MMdefMOPE}.  It is not difficult to work out the $1/\s$
corrections to the leading log asymptotics, see \cite{Basso:2021omx}.

\subsection{ Solution in regime II: Square fishnet with $ \sigma =0$
and $\varphi  \sim m$ } \la{sec:squa}
       
Before giving the general solution, it is instructive to consider a
special case which in spite of its simplicity exhibits the main new
features in the regime II. This is the case $\ell=\s=0$ which
corresponds to a square fishnet with a kinematical constraint
$x_{12}^2 x_{34}^2=x_{14}^2 x_{23}^2$.

The solution depends on the remaining parameter $\varphi  $ and connects
analytically the bulk thermodynamical limit ($\varphi =0$) and the double
light-like limit ($\varphi \to\infty$) of the square fishnet.  Because the
derivative of the external potential in this case is piecewise
constant,
\be V'(t) = \th\left( |\t |- |\varphi |\right) \, \mathrm{sgn}  \left( t\right), \ee
all the integrals evaluate to elementary functions.  The equations
\re{asymcds} for the two branch points give $b=0$ and $a = \sqrt{\varphi 
^2+4 \pi ^2 m^2}$, and the function \re{origint} reads
 \bee \H (t)&= \frac{i}{2 \pi }\, \, \mathrm{sgn}  \left( \Re t\right) \log
 \left(\frac{\sqrt{a^2-\varphi  ^2}+i \sqrt{t^2-a^2}}{\sqrt{a^2-\varphi  ^2}-i
 \sqrt{t^2-a^2}}\right) +\frac{1}{2} \, \, \mathrm{sgn}  \left( \Re t\right) \theta\left(
 |\varphi  | - |\Re t| \right)
.
\eee
  The saddle-point density
\bee \la{rhosqu} \rho(t)
&= {1\over 2 \pi^2} \log \left| \frac{\sqrt{\varphi  ^2+4 \pi ^2 m^2-t^2}+2
\pi m}{\sqrt{\varphi  ^2+4 \pi ^2 m^2-t^2}-2 \pi m}\right| , \quad 0=b\le
|t | \le a=\sqrt{\varphi  ^2+4 \pi ^2 m^2} \la{rsquare} ,\eee
 continued by symmetry to negative $t$ has a profile shown in fig.
 \ref{fig:HImRe}.  It exhibits a logarithmic cusp localised at $t=\varphi 
 $ which is a consequence of the non-analyticity of the external
 potential at this point at scale $t\sim m$.  Near the cusp the
 density behaves as
\be \rho(t) _{\mathrm{sing}} \approx {1\over 2\pi ^2 } \log|t-\varphi  |+
\text{smooth function}, \qquad 1\ll |t-\varphi  |\ll m .  \ee

\begin{figure}[tbp]
\centering 
\includegraphics[width=0.45\textwidth]{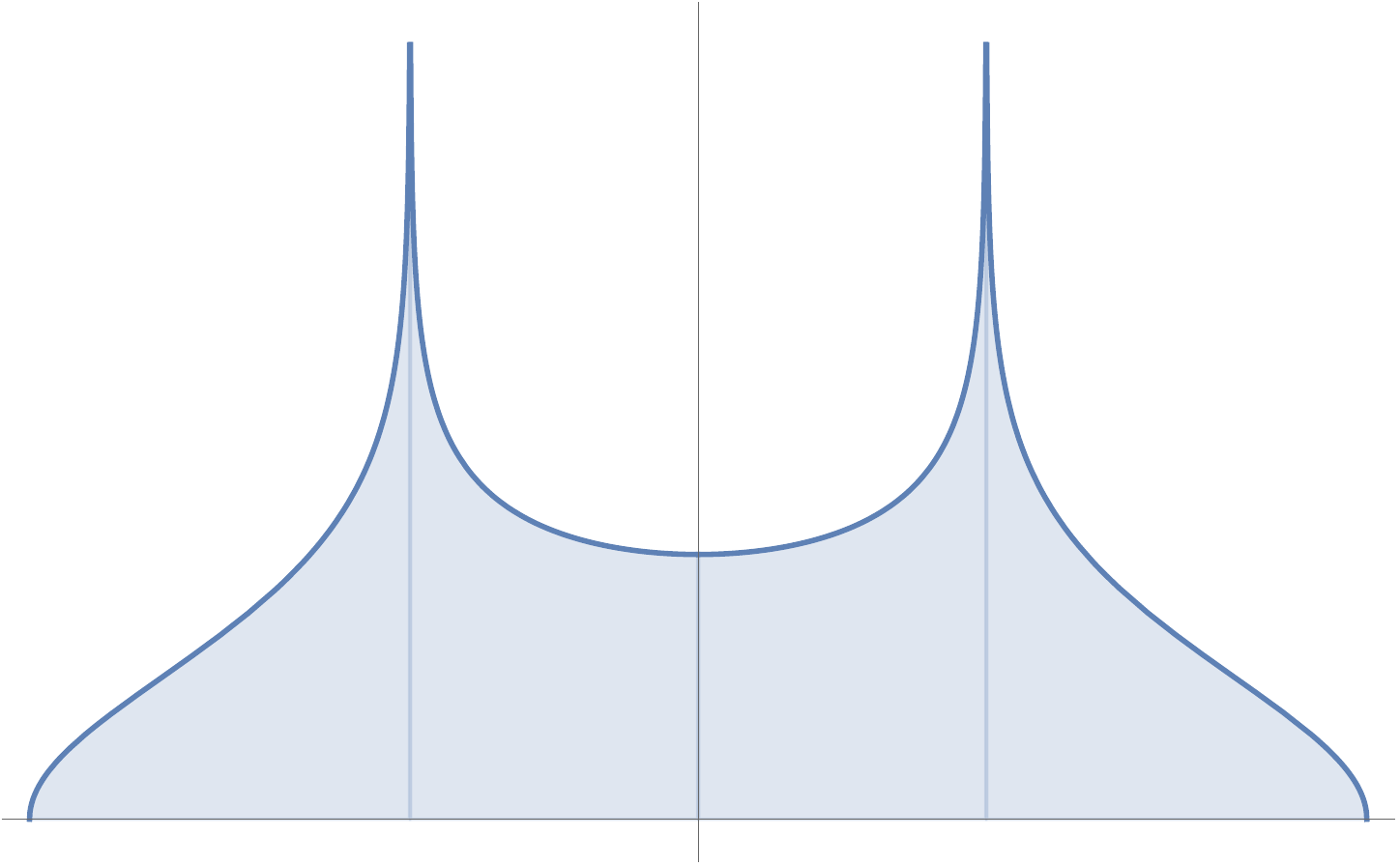}
  \caption{\small Profile of the spectral density for a large square
  fishnet with $\sigma =0$.  The density is finite at $t=0$ and
  develops a cusp at $t =\varphi $.  }
\label{fig:HImRe}
\end{figure}

 In this simple case the expression for the derivative of the  free energy
 \bee \la{dmFLL} \partial_m{\mathcal{F}} &= 2 m \log
 \left(\frac{\varphi ^2+4 \pi ^2 m^2}{16 m^2}\right) +\frac{2 \varphi
 }{\pi } \mathrm{arccot} \left(\frac{ \varphi }{ 2\pi m}\right)-\log
 (2 \pi m)+O(1) \eee
can be integrated  explicitly, with the integration constant 
fixed by the condition that the
free energy vanishes at $m=0$,
\bee \la{FRE0eta} {\mathcal{F}}&= m^2 \log \left(\frac{\varphi  ^2+4 \pi ^2 m^2}{16
m^2}\right)-\frac{\varphi  ^2 }{4 \pi ^2}\log \left(\frac{\varphi  ^2+4 \pi ^2
m^2}{\varphi  ^2}\right)+\frac{2 \varphi  m }{\pi } \mathrm{arccot} \left(\frac{
\varphi  }{ 2\pi m}\right)
.
\eee

One can check that the $\varphi \gg m$ asymptotics of \re{FRE0eta}
coincides with the large $m$ asymptotics of the solution in the double
light-like limit \re{DLLscaling} with $\U=\V = e^{-|\varphi |}$,
\bee {\mathcal{F}}&= m^2 \log { \varphi  ^2\over 16 m^2} +3 m^2 + \frac{2 \pi ^2
m^4}{3 \varphi  ^2}+O(\varphi  ^{-4}).  \eee

\begin{figure}[tbp]
\centering
\includegraphics[width=0.30\textwidth]{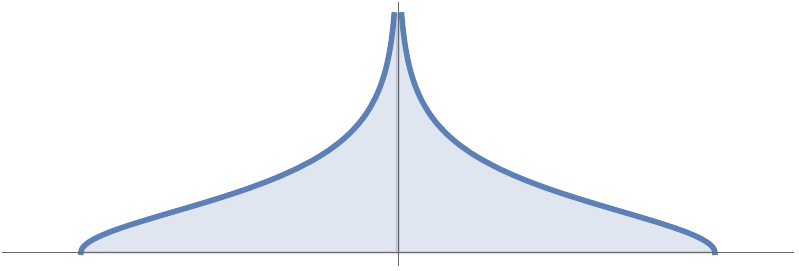}\hskip 2cm
\includegraphics[width=0.25\textwidth]{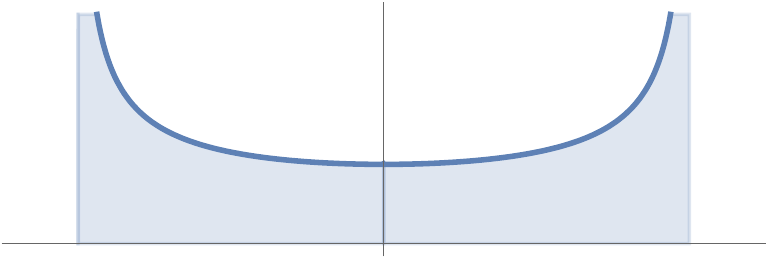}
\caption{\small When $\varphi \to 0$, the cusp moves to the origin and the
density becomes singular at $t=0$ (left).  When $ \varphi  \to\infty$, the
(right).  }
\label{fig:HImReC}
\end{figure}

\noindent The solution \re{FRE0eta} interpolates smoothly between the
bulk thermodynamical limit $( \varphi  \to 0)$ and the double light-like
limit ($ \varphi  \to\infty$) of the large square fishnet.  The first term
in the small $ \varphi  $ expansion
\bee {\mathcal{F}}&= m^2 \log{\pi^2\over 4}+m \varphi + ( \log
{\varphi ^2 \over 4 \pi^2 m^2}-3) {\varphi ^2\over 4\pi^2} +O(\varphi
^4) \eee
 matches the value $\log(\pi^2/4)$ for the free energy density,
 computed in \cite{Basso:2021omx}, as it should.  The typical shape of
 the spectral density in the two limits is shown in fig.
 \ref{fig:HImRe}.  Curiously, the expansion coefficients for large
 $\varphi $ and for small $\varphi $ are almost identical after an
 appropriate rescaling.  A similar phenomenon has been observed for
 the weak/strong coupling expansions of the dressing phase in
 ${\mathcal{N}}=4$ SYM \cite{BES}.  Here there is a simple technical
 explanation this phenomenon.  It is easy to check that the
 free-energy density as a function of the scaling variable $\hat
 \varphi $ normalised as in \re{doublelight},
\be \hat {\mathcal{F}}(\hat\varphi ) ={{\mathcal{F}}\over m^2} ,
\qquad \hat \varphi = {\varphi \over 2\pi m}, \ee
 transform under inversion $\hat\varphi \to 1/\hat\varphi $ in a very
 simple way, namely
\be {\hat F(\hat \varphi ) - \log{\pi^2\over 4}\over \hat \varphi } +
{\hat F(\hat \varphi ^{-1}) - \log{\pi^2\over 4}\over \hat \varphi
^{-1}} = 2\pi.  \ee

 \subsection{ Solution in the regime II: the general case}

Now let us consider the most general case with all the three
parameters $\ell, \sigma , \varphi $ scaling linearly with $m$.  With
$V'(t)$ given by \re{LeadingPot}, the meromorphic function
\re{origint} and the equations for the branch points are expressed in
terms of incomplete elliptic integrals (see appendix
\ref{app:Elliptic})
 \begin{align}
\la{ResolventGen} \H (t) &= \frac{t \ell }{t^2-\sigma ^2}\frac{\
\sqrt{a^2-t^2} \sqrt{b^2-t^2} }{\sqrt{a^2-\sigma ^2} \sqrt{b^2-\sigma
^2}} + {t \over \pi a } \ {\sqrt{t^2-b^2 } \over \sqrt{ t^2-a^2} } \
\Pi \left(\frac{a^2-b^2}{a^2-t^2};\psi \Big| k^2\right), \\ \no \\
  & {F\left( \psi\big| k^2\right) } \, \sqrt{ \left( a^2
  -\sigma^2\right) \left(b ^2 -\sigma^2\right)} = \pi \ell a ,
  \la{eqfora} \\
 &  a^2 {E\left( \psi \big| k^2\right) } -\sigma ^2  F \left( \psi
  \big| k^2\right)  = \pi(2m +\ell ) a,
  \la{eqforb}
  \\
   &k^2 = 1- {b^2\over a^2}, \qquad \psi =\arcsin \frac{\sqrt{a^2-\varphi 
   ^2}}{\sqrt{a^2-b^2}} .
\end{align}
  The semiclassical spectral density
\bee \la{rhogeneral} \rho(t) = & {1 \over\pi} { \ell \, t\over
t^2-\sigma ^2} \sqrt{{(a^2-t^2)(t^2-b^2)\over (a^2-\sigma
^2)(b^2-\sigma ^2)}} + {1 \over \pi^2} {t\over a}\sqrt{t^2-b^2\over
a^2-t^2}\ \Pi\left( {a^2-b^2\over a^2-t^2}; \psi \Big| k^2\right), \eee
has a cusp at $t=\varphi $ (fig.  \ref{fig:densitygen}).  The regime I
is attained when the cusp moves to the left edge and disappears (fig.
\ref{fig:BTL-DLC}, left).  In the light-cone limit $\varphi \gg m$,
the cusp moves to the right edge and changes the square-root
singularity to a logarithmic one (fig.  \ref{fig:BTL-DLC}, right).

\begin{figure}[tbp]
\centering  
\includegraphics[width=0.65\textwidth]{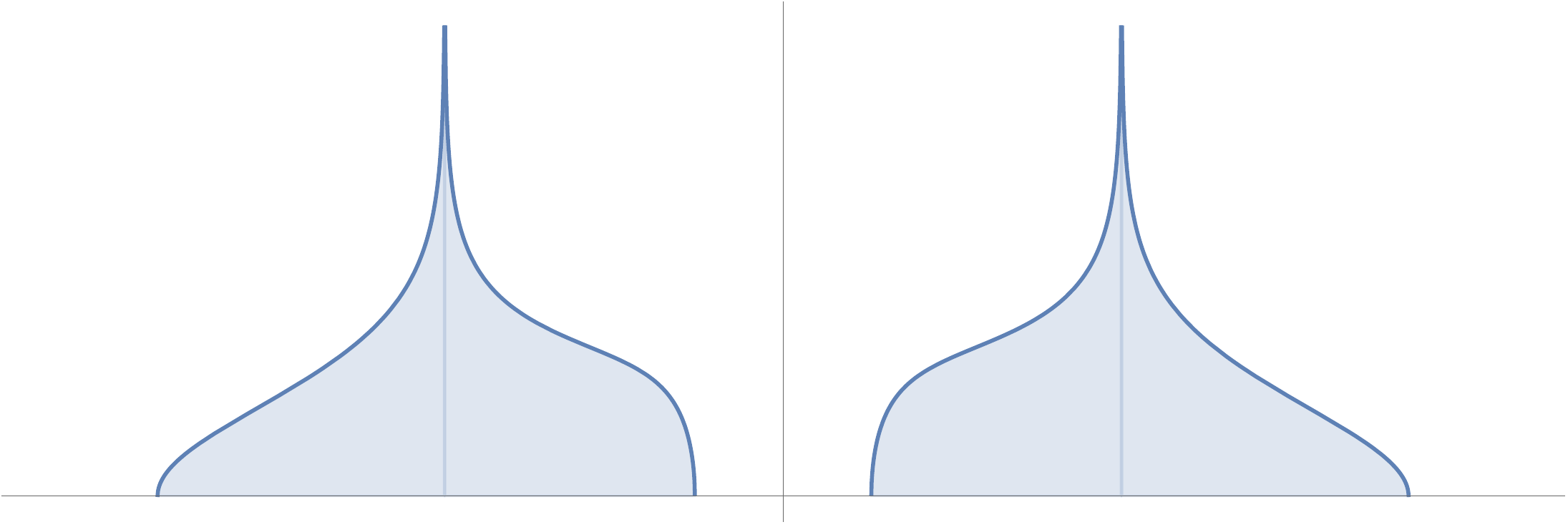}
  \hskip 2.9 cm \caption{\small Profile of the spectral density
  $\rho(t)$ in regime II. }
\label{fig:densitygen}
\end{figure}

 \begin{figure}[tbp]
\bigskip \centering
\includegraphics[width=0.35\textwidth]{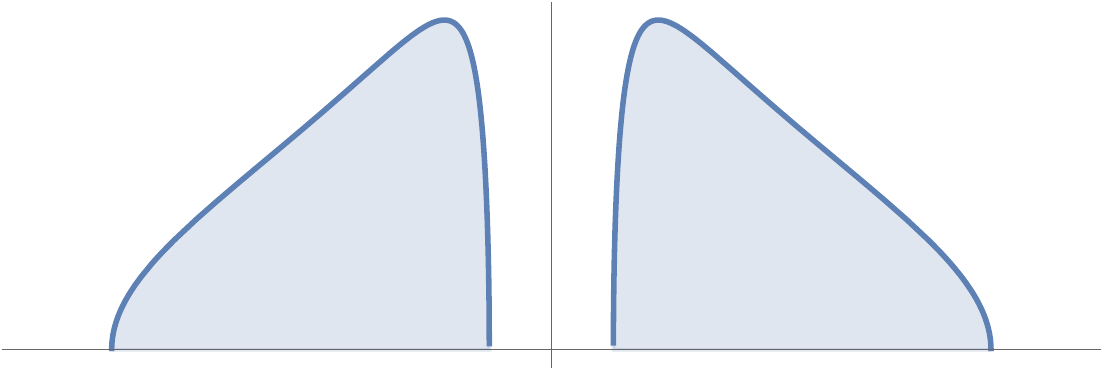}
  \hskip 2.5 cm \includegraphics[width=0.36\textwidth]{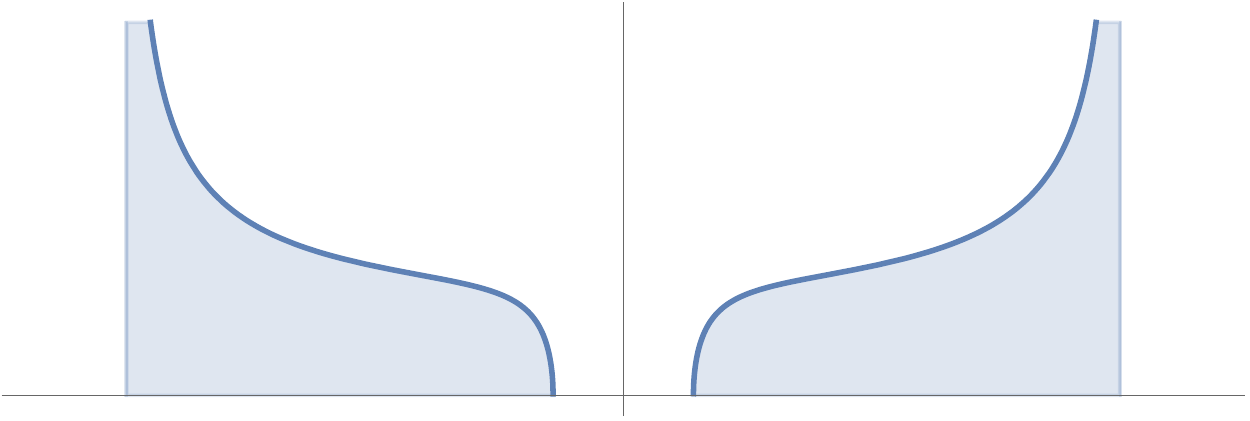}
  \caption{\small Profile of the spectral density when $b=|\varphi |$
  (left) and for $a\to |\varphi |$ (right).  }
   
\label{fig:BTL-DLC} 
\end{figure}

The expression for the derivative of the free energy,
 \bee \la{dFFFbb} \partial_m {\mathcal{F}} &= (2 m+\ell) \log{(a^2-b^2
 )\over 4 (2m+\ell)^2 } + \frac{2 \varphi }{\pi } \arctan
 \frac{\sqrt{a^2-\varphi ^2 }}{\sqrt{ \varphi ^2 -b^2}} \\
 & + 2 \ell \, \arctanh{\sqrt{b^2-\sigma ^2}\over\sqrt{a^2-\sigma ^2}}
 - \frac{ 2\ell \, \sigma ^2 }{\sqrt{ \left(a^2-\sigma ^2\right)
 \left(b^2-\sigma ^2\right) }}, \eee
is obtained by using the relations \re{eqfora} and \re{eqforb} to
express the elliptic integrals in eq.  \re{dmF} in terms of $ a$ and
$b$.
      
Eq.  \re{dFFFbb} can be used to generate series expansions of the free
energy in different limits of regime II, as the bulk thermodynmical
limit and the double light-cone limit, but this task is beyond the
scope of this paper.  Below I will only check that the limit $\s/m\gg
1$ of \re{dFFFbb} along the line $\s=0$ indeed reproduces the large
$m$ asymptotics of the expression \re{DLLscaling} for the double
light-like limit.

   \bigskip

  \noindent \bigskip $\bullet$ {\it Double light-cone limit $\sigma
  =0, \varphi  \gg m$}

If $\varphi  \gg m$, then the right branch point is pushed far as well,
$a\gg m$.  The left branch point can be anywhere depending on the
value of $\ell$.  The two conditions \re{eqfora}-\re{eqforb} are
compatible with $\psi\ll 1$.  Retaining only the leading linear order
in the expansion of the elliptic integrals in $\psi$, they read
\bee \pi {\ell\over a} = \sqrt{1-k^2} \, \psi , \ \ \ \pi {2
m+\ell\over a}= \, \psi , \ \ {\varphi  \over a}=\sqrt{1-k^2 \psi ^2},
\eee
with solution to the leading order at $\psi\ll 1$
\bee \psi& \to \frac{\pi (2 m+\ell )}{\varphi }, \ \ \ k\to \frac{2
\sqrt{m(m +\ell) }}{ 2m+\ell}, \\
 a
  &=\varphi  +\frac{2 \pi ^2 m (m+\ell )}{\varphi  },
\ \ \   
{b } = { \varphi \, \ell \over 2 m+\ell}
.
\eee
At $m\to 0$, $a=\varphi $ is the position of the minimum of the
external potential.  With the condition ${\mathcal{F}}_{m\to 0}=0$,
the derivative of the free energy can be integrated to
\bee {\mathcal{F}}&=2 m (m+\ell ) \log  \varphi  +3 m (m+\ell )\\
&+m^2 \log (m) +(m+\ell )^2 \log (m+\ell )-(2 m+\ell )^2 \log (2 m+\ell )
\\
&=2 m n \log (\varphi )+3 m n +  m^2 \log (m)+ n^2 \log
(n) -(m+n)^2 \log (m+n).  \eee
This expression matches the large-$m$ asymptotics of \re{DLLscaling}.

  \section{Discussion}
  
This short note addresses the question how the thermodynamical limit of  
 Basso-Dixon
integral for the $m\times n$
rectangular fishnets is affected by the boundary conditions.

The saddle point for the Basso-Dixon integral described by a density
function is found in the double scaling limit \re{doublelightb} where
the spacetime parameters $\s$ and $\varphi $ scale large with the
fishnet lengths $m$ and $n$.  This is the most general scaling regime
which contains the Euclidean OPE, the light-cone and the bulk
thermodynamical limits as particular cases.
 
In the double scaling limit  there are two regimes, labeled here by I and II 
and characterised by different analytic solutions.  
In regime II, a closed analytic
expression, eq.  \re{dFFFbb}, is derived for the logarithmic
derivative of the fishnet in $m$ with fixed $\ell = n-m, \s$ and
$\varphi $.  The latter determines how the fishnet changes upon
adding a new row and a new column to the rectangle grid.  With this
observation, eq.  \re{dFFFbb} can be written as \smallskip
 \bee \la{dFFFmn} &\log \left[ {I^{\mathrm{BD}}_{m+1,n+1}\over
 I^{\mathrm{BD}}_{m,n}} \right] = ( m+n) \log{ a^2 -b^2 \over 4 (
 m+n)^2 } + \frac{2 \varphi }{\pi } \arctan \frac{\sqrt{a^2-\varphi ^2
 }}{\sqrt{ \varphi ^2 -b^2 }} \\
 & + 2 |n-m| \left( \arctanh{\sqrt{ b^2-\sigma ^2}\over\sqrt{a^2-\sigma
 ^2}} - \frac{ \sigma ^2 }{\sqrt{ \left(a^2-\sigma ^2\right) \left(
 b^2-\sigma ^2\right) }} \right), \\
& {F\left( \psi\big| k^2\right) }
  \, \sqrt{ \left( a^2 -\sigma^2\right) \left( b^2-\sigma^2\right)}
 = \pi    a  |n-m|, 
 \\
 & a^2 {E\left( \psi \big| k^2\right) } -\sigma ^2 F \left( \psi \big| k^2\right) = \pi
 a ( m + n ), \\
  & k^2 = 1- {b^2\over a^2}, \ \   \psi =\arcsin
  {\sqrt{a^2-\varphi  ^2} \over \sqrt{a^2-b^2}} 
 \qquad \qquad(m,n\gg 1; \varphi \ge  b\le  \s\sim m) .
  \eee
Although $m$ and $n$ enter in a non-symmetric way in the original
integral, the result is symmetric under $m\leftrightarrow n$.  A
direct integration is possible only in some cases as e.g. for
$\s=\ell=0$ where the solution is given in a closed form in section
\ref{sec:squa}.  The solution for the regime I is obtained by taking
$\psi = \pi/2$ in \re{dFFFmn}.

The saddle-point equations for the large $m$ limit of the Basso-Dixon
integral take the form of the Bethe equations for a generalised
$sl(2)$ spin chain, with the role of the BMN coupling constant played
by the parameter $\s$.  The second parameter $\varphi $ appears only after
the Bethe equations are written in a logarithmic form.

 In the regime I which has been investigated in \cite{Basso:2021omx},
 the equation for the density of the Bethe roots is a finite gap
 equation for a symmetric configuration where all positive Bethe roots
 have mode number $ 1$.  The solution describes the Frolov-Tseytlin
 folded string rotating in $AdS_3\times S^1$.  In regime II, the
 solution of the Bethe equations exhibits a completely new feature.
 The positive roots split into two groups with mode numbers $1$ and
 $0$.  The roots larger than $\varphi $ have mode number 1 while the
 root smaller that $\varphi $ have mode number 0.  Such a choice of
 the mode numbers is mathematically possible, but does not lead to a
 finite-gap solution because the two groups of Bethe roots do not
 repel but attract.  As a consequence, the eigenvalue density develops
 a logarithmic cusp at the point $\varphi $.  Such a solution does not
 seem to correspond to any string motion in AdS.

 The dual AdS description of the open fishnets thus remains an open
 problem.  Looking ahead, one possible direction is to try to adjust
 the non-linear sigma model \cite{Basso:2018agi} or the string-bit
 \cite{Gromov:2019aku,Gromov:2019bsj} formulations of the cylindrical
 fishnet.  Another direction would be to connect to the holographic
 description of the octagon \cite{Coronado:2018cxj} for which
 fascinating exact results have been obtained recently
 \cite{Belitsky:2019fan,Belitsky:2020qrm,Belitsky:2020qir,
 Belitsky:2020qzm,Bargheer:2019kxb,Bargheer:2019exp}, using the fact
 that the fishnet is obtained from the octagon by truncation.  A more
 relevant quantity in this respect would be the grand partition
 function for rectangular fishnet obtained as discrete Laplace
 transform with respect to the fishnet lengths $m$ and $n$.
 
Another new in my knowledge result concerns the double light-cone
limit with the four points approaching the cusps of a null square,
where an exact expression of the rectangular fishnet is found for any
$m$ and $n$.  This expression, eq.  \re{DLLscaling}, factorises into
two pieces associated with the two cross ratios which become singular.
The numerical factor in each piece is the same as the one obtained in
\cite{Basso:2021omx} for the Euclidean short-distance limit.  The
origin of this factorisation is yet to be understood.  In the case of
the Euclidean OPE limit, a combinatorial interpretation of the leading
log asymptotics \re{MMdefMOPE} was recently given in
\cite{Olivucci:2021pss}.  The rectangular fishnet was interpreted as
an amplitude for hopping magnons named `stampede'.\footnote{Curiously,
a similar combinatorics occurs in the evaluation of the Jordan block
spectrum of the `hypereclectic spin chain'
\cite{Ahn:2021emp,Ahn:2022snr}.} The similarity of \re{MMdefMOPE} and
\re{DLLscaling} suggests that a description in terms of hopping
magnons is possible also in the double light-cone limit.

The `stampede' interpretation could unveil interesting physics.  For
example, the conjecture made in \cite{Basso:2021omx} about the
possibility of arctic-curve phenomenon like in the six-vertex model
with domain-wall boundary conditions \cite{2005cond.mat..2314A}, can
be given more concrete shape here.  Namely it is very likely that the
typical `stampede' exhibits in the thermodynamical limit fluctuating
and frozen phases separated by an `arctic curve'.

\acknowledgments

I am indebted to Benjamin Basso, Nikolay Gromov and Enrico Olivucci
for several interesting and useful discussions.  This research was
supported in part by the National Science Foundation under Grant No.
NSF PHY-1748958

   \appendix

   \section{Elliptic integrals}
   \la{app:Elliptic}
   
$ \K= {\tt EllipticF[ \psi, k^2]}$ and $\mathbb{E}= {\tt EllipticE[ \psi,
k^2]}$ are respectively the complete elliptic integrals of first and
second kind,
   \bee \K=\int_0^{\pi/2} {d\theta\over \sqrt{1-k^2 \sin^2 \theta }},
   \quad \E&=\int_0^{\pi/2} d\theta\, \sqrt{1-k^2 \sin^2 \theta } ,
   \eee
 $\Pi(\a^2|k^2)= {\tt EllipticPi} [\a^2,k^2] $ is the complete
 elliptic integral of third kind,
  \bee \Pi(\a^2|k^2)&= \int_0^{\pi/2} {d \theta \over (1- \a^2
  \sin^2\th)\sqrt{1- k^2 \sin^2\theta}}.  \eee
	$ \Lambda_0(\psi , k^2) $ denotes Heuman's Lambda function,
 \bee \Lambda_0(\psi , k^2) &\equiv {2\over \pi} \left[  \mathbb{E} \, F(\psi , k')
 + \mathbb{K} \, E(\psi , k') - \K\, F(\psi , k')\right] .  \eee

$ F(\psi| k^2)= {\tt EllipticF[ \psi, k^2]}$ and $E(\psi| k^2)= {\tt
EllipticE[ \psi, k^2]}$ are respectively the incomplete elliptic
integrals of first and second kind,
  \bee F (\psi |k^2 )&=\int_0^{\psi } {d\theta\over \sqrt{1-k^2 \sin^2
  \theta }} , \quad E (\psi |k^2 )&=\int_0^{\psi } d\theta\,
  \sqrt{1-k^2 \sin^2 \theta } .  \eee

$\Pi(\a^2; \psi|k^2)= {\tt EllipticPi} [\a^2,\psi,k^2] $ is the
incomplete elliptic integral of third kind;
\bee \Pi(\a^2; \psi|k^2)&=
\int_0^\psi {d \theta \over (1- \a^2 \sin^2\th)\sqrt{1- k^2
\sin^2\theta}}.  \eee

\section{Fermionic representation of the $(m+\ell)\times m$
rectangular fishnet and derivation of the dual integral} \la{app:FF}

 Consider the Fock space of a complex fermion
   \bee\la{defcferm} &\psi (u )= \sum_{n\ge 0} \psi _{n } \ u ^{ -n-1
   },\quad \psi^* (u) = \sum_{n\ge 0} \psi^ * _{n} \ u ^{ n} , \qquad
   [\psi_m, \psi^ * _n]_+ =\delta_{m,n} \eee
 with vacuum states defined by $\langle 0 | \psi^ * _{n} =0$ and $
 \psi_n | \ell \rangle = 0$.  The vacuum states of charge $\ell$ are
 constructed as $ \< \ell| = \< 0| \psi_0\psi_1...\psi_\ell $ and $
 |\ell\> = \psi^*_\ell...\psi^*_0 |0\>$.
The statement is that the Basso-Dixon integral \re{fishnetintegral}
equals the matrix element
\be \la{FockexpV} {I^{\mathrm{BD}} _{m,n} }= \<\ell | \ e^{ \mathbf{H} } \
|\ell+ 2 m \> \ee
of the evolution operator is constructed from the fermion bilinear
  \bee \la{FishBMNFF} \mathbf{H} &={1\over \sqrt{\U\V}}
   \sum_{a \ge 1}  { \sin a\varphi 
  \over \sin\varphi  } \int {du \over 2\pi } \ \ {e^{2 i \sigma u} } \ \psi
  (u + ia/2)\psi (u - ia/2) \\
&= \int _{{\mathbb{R}} + i 0} {du \over 2\pi i} \psi (u+i0 ) \ e^{ 2 i
\sigma u} \ { \cosh \sigma + \cosh \varphi \over \cos\partial_u +
\cosh \varphi } \ \ \psi (u-i0 ) .  \eee
The fermionic representation \re{FockexpV}, which is similar to that
of the octagon \cite{Kostov:2021omc}, can be proved by expanding the
exponent and using the expression for the two-point function
 \be \<\ell| \psi(u)\psi(v)|\ell+2\> = u^{-\ell -1} v^{-\ell - 2} -
 v^{-\ell -1} u^{-\ell - 2} ={u-v\over (u v)^{2+\ell}}.  \ee
The second line in \re{FishBMNFF} is a formal expression for the sum
in the first line.  It can be given precise meaning by honestly
performing Fourier transformation and using the identity
\be \la{defX} ({\U \V })^{-1/2} \sum _{a\ge 1} e^{- a t}\ {\sinh
a\varphi \over \sinh\varphi } = { \cosh\sigma +\cosh\varphi \over
\cosh t +\cosh\varphi } \equiv e^{-V_0(t)} .  \ee
The Fourier transformed fermion is given by two different expressions
depending on whether the argument is below or above the real axis:
 \be \psi (u \pm i 0 ) = \int_{-\infty}^\infty {dt }\, e^{- i u t}\
 \tilde \psi _\pm (t), \ \ \ \ \tilde \psi _\pm (t) =
 \int_{-\infty}^\infty {du\over 2\pi}\, e^{ i u t} \psi (u \pm i 0) .
 \ee
 or in terms of the mode expansion \re{defcferm}, \be \psi (u \pm i
 0)= \sum_{n\ge 0} \psi _{n } \ (u \pm i 0)^{ -n-1}, \quad \tilde \psi
 _\pm(t) = \mp i \th(\mp t) \sum_{n\ge0} \psi _{n } \ {(it)^{ n }\over
 n!}.  \ee
 The two-point function of the Fourier-transformed fermion
 $\tilde\psi_\pm$ is
\be \begin{aligned}
  \< \ell|\tilde \psi_+(t_1) \tilde \psi_-(t_2) |\ell+2 \>&= -
  {i^{2\ell+1}\over \ell!(\ell+1)!}\ t_1 ^\ell t_2 ^\ell \ (t_2-t_1)\
  \th (-t_1)\th(t_2)\\
  \< \ell|\tilde \psi_-(t_1) \tilde \psi_-(t_2)
 |\ell+2 \>&= - {i^{2\ell+1}\over \ell!(\ell+1)!}\   t_1 ^\ell  t_2 ^\ell
 \
 (t_2-t_1)\ \th (t_1)\th(t_2), \quad \text{etc}.
 \end{aligned} \ee
The shifts in $\pm i a/2$ acts in the Fourier space diagonally while
the factor $e^{i\sigma u}$ transforms into a shift $t\to t+\sigma $,
 \bee e^{i\sigma u} \psi (u \pm {\textstyle{i\over 2} } a ) &=
 \int_{-\infty}^\infty } {dt\, \theta(\mp t\mp\s) \, e^{\pm a t/2}\,
 e^{-i u t} \tilde \psi _\pm (t+\sigma ) \\
  &= \mp  \int_{  \mp \sigma }^\infty {dt
 }\,  e^{-a t/2}\,   e^{\pm i u t} \tilde \psi _\pm (\sigma \mp t),  
 \eee
where we have used that $\tilde \psi_\pm (t) = \theta (\mp t)\tilde
\psi_\pm (t) $.  Therefore
\be\begin{split} \la{Fourierpairs} \int {du\over 2\pi}\ e^{2 i \sigma
u} \ \psi (u + {\textstyle{i\over 2} } a) \psi (u - {\textstyle{i\over
2} } a ) &= -\int_{|\sigma |}^\infty {dt }\, \tilde \psi _+(\sigma
-t)\ \tilde \psi _-(\sigma +t) \ e^{-at} \end{split}.  \ee
The weighted sum in $a\ge 1$ can be performed in the exponent,
producing
  \begin{align}
\tilde{ \mathbf{H}} =- \int_{|\sigma |}^{ \infty} dt\, e^{-V_0(t)} \
\tilde \psi _+(\sigma -t)\ \tilde \psi _-(\sigma +t),
   \end{align}
   with $V_0(t)$  defined  in \re{defX}.
% %
Evaluating the expectation value by performing all Wick contractions,
one obtains the dual representation of the rectangular fishnet in
the Fourier space for the rapidities, eq.  \re{MMpartf},
\bee\begin{aligned} {I^{\mathrm{BD}} _{m,n} } & = \<\ell | \ e^{ \tilde{
\mathbf{H}} } \ |\ell+ 2 m \> \\ & = \prod_{j=1}^m \int_{|\sigma
|}^\infty dt_j\ { e^{-V_0(t_j)} (\sigma ^2-t_j^2)^\ell \over
(\ell+2j-2)!  (\ell+2j-1)!} \ \prod_{j,k =1}^m (t_j+t_k) \prod_{j<k}^m
(t_j-t_k)^2 .  \end{aligned}\eee
The determinant representation \re{detrep} follows from the expression
of the expectation value as a $2m\times 2m$ pfaffian, which can be
turned into a determinant.

%
%  \bibliography{/Users/vani/Dropbox/ABib-dropbox-last.bib}

\begin{thebibliography}{10}

\bibitem{USSYUKINA1993363}
N.~Ussyukina and A.~Davydychev, \emph{An approach to the evaluation of three-
  and four-point ladder diagrams},
  \href{https://doi.org/https://doi.org/10.1016/0370-2693(93)91834-A}{\emph{Physics
  Letters B} {\bfseries 298} (1993) 363 }.

\bibitem{Zamolodchikov:1980mb}
A.B.~Zamolodchikov, \emph{{'Fishnet' diagrams as a completely integrable
  system}}, {\emph{Phys. Lett.} {\bfseries B97} (1980) 63}.

\bibitem{Gurdogan:2015csr}
{\"O}.~G{\"u}rdo{\u g}an and V.~Kazakov, \emph{{New Integrable 4D Quantum Field
  Theories from Strongly Deformed Planar $\mathcal N = $ 4 Supersymmetric
  Yang-Mills Theory}}, \href{https://doi.org/10.1103/PhysRevLett.117.201602,
  10.1103/PhysRevLett.117.259903}{\emph{Phys. Rev. Lett.} {\bfseries 117}
  (2016) 201602} [\href{https://arxiv.org/abs/1512.06704}{{\ttfamily
  1512.06704}}].

\bibitem{Caetano:2016ydc}
J.a.~Caetano, O.~G\"urdo\u{g}an and V.~Kazakov, \emph{{Chiral limit of $
  \mathcal{N} $ = 4 SYM and ABJM and integrable Feynman graphs}},
  \href{https://doi.org/10.1007/JHEP03(2018)077}{\emph{JHEP} {\bfseries 03}
  (2018) 077} [\href{https://arxiv.org/abs/1612.05895}{{\ttfamily
  1612.05895}}].

\bibitem{Grabner:2017pgm}
D.~Grabner, N.~Gromov, V.~Kazakov and G.~Korchemsky, \emph{{Strongly
  $\gamma$-Deformed $\mathcal{N}=4$ Supersymmetric Yang-Mills Theory as an
  Integrable Conformal Field Theory}},
  \href{https://doi.org/10.1103/PhysRevLett.120.111601}{\emph{Phys. Rev. Lett.}
  {\bfseries 120} (2018) 111601}
  [\href{https://arxiv.org/abs/1711.04786}{{\ttfamily 1711.04786}}].

\bibitem{Gromov:2017cja}
N.~Gromov, V.~Kazakov, G.~Korchemsky, S.~Negro and G.~Sizov,
  \emph{{Integrability of Conformal Fishnet Theory}},
  \href{https://doi.org/10.1007/JHEP01(2018)095}{\emph{JHEP} {\bfseries 01}
  (2018) 095} [\href{https://arxiv.org/abs/1706.04167}{{\ttfamily
  1706.04167}}].

\bibitem{Basso:2018agi}
B.~Basso and D.-l.~Zhong, \emph{{Continuum limit of fishnet graphs and AdS
  sigma model}}, \href{https://doi.org/10.1007/JHEP01(2019)002}{\emph{JHEP}
  {\bfseries 01} (2019) 002}
  [\href{https://arxiv.org/abs/1806.04105}{{\ttfamily 1806.04105}}].

\bibitem{Gromov:2019bsj}
N.~Gromov and A.~Sever, \emph{{Quantum fishchain in AdS$_{5}$}},
  \href{https://doi.org/10.1007/JHEP10(2019)085}{\emph{JHEP} {\bfseries 10}
  (2019) 085} [\href{https://arxiv.org/abs/1907.01001}{{\ttfamily
  1907.01001}}].

\bibitem{Gromov:2019aku}
N.~Gromov and A.~Sever, \emph{{Derivation of the Holographic Dual of a Planar
  Conformal Field Theory in 4D}},
  \href{https://doi.org/10.1103/PhysRevLett.123.081602}{\emph{Phys. Rev. Lett.}
  {\bfseries 123} (2019) 081602}
  [\href{https://arxiv.org/abs/1903.10508}{{\ttfamily 1903.10508}}].

\bibitem{Gromov:2019jfh}
N.~Gromov and A.~Sever, \emph{{The Holographic Dual of Strongly
  $\gamma$-deformed N=4 SYM Theory: Derivation, Generalization, Integrability
  and Discrete Reparametrization Symmetry}},
  \href{https://doi.org/10.1007/JHEP02(2020)035}{\emph{JHEP} {\bfseries 02}
  (2020) 035} [\href{https://arxiv.org/abs/1908.10379}{{\ttfamily
  1908.10379}}].

\bibitem{Basso:2019xay}
B.~Basso, G.~Ferrando, V.~Kazakov and D.-l.~Zhong, \emph{{Thermodynamic Bethe
  Ansatz for Biscalar Conformal Field Theories in any Dimension}},
  \href{https://doi.org/10.1103/PhysRevLett.125.091601}{\emph{Phys. Rev. Lett.}
  {\bfseries 125} (2020) 091601}
  [\href{https://arxiv.org/abs/1911.10213}{{\ttfamily 1911.10213}}].

\bibitem{Drummond:Yangian}
J.~Drummond, J.~Henn and J.~Plefka, \emph{Yangian symmetry of scattering
  amplitudes in n=4 super yang-mills theory}, {\emph{JHEP} {\bfseries 0905}
  (2009) 046} [\href{https://arxiv.org/abs/0902.2987}{{\ttfamily 0902.2987}}].

\bibitem{Frassek:2013xza}
R.~Frassek, N.~Kanning, Y.~Ko and M.~Staudacher, \emph{{Bethe Ansatz for
  Yangian Invariants: Towards Super Yang-Mills Scattering Amplitudes}},
  \href{https://doi.org/10.1016/j.nuclphysb.2014.03.015}{\emph{Nucl. Phys. B}
  {\bfseries 883} (2014) 373}
  [\href{https://arxiv.org/abs/1312.1693}{{\ttfamily 1312.1693}}].

\bibitem{Chicherin:2017frs}
D.~Chicherin, V.~Kazakov, F.~Loebbert, D.~M\"uller and D.-l.~Zhong,
  \emph{{Yangian Symmetry for Fishnet Feynman Graphs}},
  \href{https://doi.org/10.1103/PhysRevD.96.121901}{\emph{Phys. Rev. D}
  {\bfseries 96} (2017) 121901}
  [\href{https://arxiv.org/abs/1708.00007}{{\ttfamily 1708.00007}}].

\bibitem{Chicherin:2022nqq}
D.~Chicherin and G.P.~Korchemsky, \emph{{The SAGEX Review on Scattering
  Amplitudes, Chapter 9: Integrability of Amplitudes in Fishnet Theories}},
  \href{https://arxiv.org/abs/2203.13020}{{\ttfamily 2203.13020}}.

\bibitem{Basso:2021omx}
B.~Basso, L.J.~Dixon, D.A.~Kosower, A.~Krajenbrink and D.-l.~Zhong,
  \emph{{Fishnet four-point integrals: integrable representations and
  thermodynamic limits}},
  \href{https://doi.org/10.1007/JHEP07(2021)168}{\emph{JHEP} {\bfseries 07}
  (2021) 168} [\href{https://arxiv.org/abs/2105.10514}{{\ttfamily
  2105.10514}}].

\bibitem{Basso:2017jwq}
B.~Basso and L.J.~Dixon, \emph{{Gluing Ladder Feynman Diagrams into Fishnets}},
  \href{https://doi.org/10.1103/PhysRevLett.119.071601}{\emph{Phys. Rev. Lett.}
  {\bfseries 119} (2017) 071601}
  [\href{https://arxiv.org/abs/1705.03545}{{\ttfamily 1705.03545}}].

\bibitem{Berenstein:2002jq}
D.E.~Berenstein, J.M.~Maldacena and H.S.~Nastase, \emph{{Strings in flat space
  and pp waves from N=4 superYang-Mills}},
  \href{https://doi.org/10.1088/1126-6708/2002/04/013}{\emph{JHEP} {\bfseries
  0204} (2002) 013} [\href{https://arxiv.org/abs/hep-th/0202021}{{\ttfamily
  hep-th/0202021}}].

\bibitem{BKV1}
B.~Basso, S.~Komatsu and P.~Vieira, \emph{{Structure Constants and Integrable
  Bootstrap in Planar N=4 SYM Theory}},
  \href{https://arxiv.org/abs/1505.06745}{{\ttfamily 1505.06745}}.

\bibitem{Fleury:2016ykk}
T.~Fleury and S.~Komatsu, \emph{{Hexagonalization of Correlation Functions}},
  \href{https://doi.org/10.1007/JHEP01(2017)130}{\emph{JHEP} {\bfseries 01}
  (2017) 130} [\href{https://arxiv.org/abs/hep-th/1611.05577}{{\ttfamily
  hep-th/1611.05577}}].

\bibitem{Eden:2016xvg}
B.~Eden and A.~Sfondrini, \emph{{Tessellating cushions: four-point functions in
  $\mathcal{N} $ = 4 SYM}},
  \href{https://doi.org/10.1007/JHEP10(2017)098}{\emph{JHEP} {\bfseries 10}
  (2017) 098} [\href{https://arxiv.org/abs/hep-th/1611.05436}{{\ttfamily
  hep-th/1611.05436}}].

\bibitem{Fleury:2017eph}
T.~Fleury and S.~Komatsu, \emph{{Hexagonalization of Correlation Functions II:
  Two-Particle Contributions}},
  \href{https://doi.org/10.1007/JHEP02(2018)177}{\emph{JHEP} {\bfseries 02}
  (2018) 177} [\href{https://arxiv.org/abs/hep-th/1711.05327}{{\ttfamily
  hep-th/1711.05327}}].

\bibitem{Basso:2018cvy}
B.~Basso, J.a.~Caetano and T.~Fleury, \emph{{Hexagons and Correlators in the
  Fishnet Theory}}, \href{https://doi.org/10.1007/JHEP11(2019)172}{\emph{JHEP}
  {\bfseries 11} (2019) 172}
  [\href{https://arxiv.org/abs/1812.09794}{{\ttfamily 1812.09794}}].

\bibitem{BSV-1}
B.~Basso, A.~Sever and P.~Vieira, \emph{Spacetime and flux tube s-matrices at
  finite coupling for n=4 supersymmetric yang-mills theory},
  \href{https://doi.org/10.1103/PhysRevLett.111.091602}{\emph{Phys. Rev. Lett.}
  {\bfseries 111} (2013) 091602}.

\bibitem{BSV-2}
B.~Basso, A.~Sever and P.~Vieira, \emph{{Space-time S-matrix and Flux tube
  S-matrix II. Extracting and Matching Data}},
  \href{https://doi.org/10.1007/JHEP01(2014)008}{\emph{JHEP} {\bfseries 01}
  (2014) 008} [\href{https://arxiv.org/abs/1306.2058}{{\ttfamily 1306.2058}}].

\bibitem{Derkachov:2020zvv}
S.~Derkachov and E.~Olivucci, \emph{{Exactly solvable single-trace four point
  correlators in $\chi$CFT$_4$}},
  \href{https://doi.org/10.1007/JHEP02(2021)146}{\emph{JHEP} {\bfseries 02}
  (2021) 146} [\href{https://arxiv.org/abs/2007.15049}{{\ttfamily
  2007.15049}}].

\bibitem{Derkachov:2021rrf}
S.~Derkachov and E.~Olivucci, \emph{{Conformal quantum mechanics \& the
  integrable spinning Fishnet}},
  \href{https://doi.org/10.1007/JHEP11(2021)060}{\emph{JHEP} {\bfseries 11}
  (2021) 060} [\href{https://arxiv.org/abs/2103.01940}{{\ttfamily
  2103.01940}}].

\bibitem{Chicherin:2012yn}
D.~Chicherin, S.~Derkachov and A.P.~Isaev, \emph{{Conformal group: R-matrix and
  star-triangle relation}},
  \href{https://doi.org/10.1007/JHEP04(2013)020}{\emph{JHEP} {\bfseries 04}
  (2013) 020} [\href{https://arxiv.org/abs/1206.4150}{{\ttfamily 1206.4150}}].

\bibitem{Derkachov:2018rot}
S.~Derkachov, V.~Kazakov and E.~Olivucci, \emph{{Basso-Dixon Correlators in
  Two-Dimensional Fishnet CFT}},
  \href{https://doi.org/10.1007/JHEP04(2019)032}{\emph{JHEP} {\bfseries 04}
  (2019) 032} [\href{https://arxiv.org/abs/1811.10623}{{\ttfamily
  1811.10623}}].

\bibitem{Broadhurst:2010ds}
D.J.~Broadhurst and A.I.~Davydychev, \emph{{Exponential suppression with four
  legs and an infinity of loops}},
  \href{https://doi.org/10.1016/j.nuclphysbps.2010.09.014}{\emph{Nucl. Phys.
  Proc. Suppl.} {\bfseries 205-206} (2010) 326}
  [\href{https://arxiv.org/abs/1007.0237}{{\ttfamily 1007.0237}}].

\bibitem{2000JPhA...33.7053K}
V.~{Korepin} and P.~{Zinn-Justin}, \emph{{Thermodynamic limit of the six-vertex
  model with domain wall boundary conditions}},
  \href{https://doi.org/10.1088/0305-4470/33/40/304}{\emph{Journal of Physics A
  Mathematical General} {\bfseries 33} (2000) 7053}
  [\href{https://arxiv.org/abs/arXiv:cond-mat/0004250}{{\ttfamily
  arXiv:cond-mat/0004250}}].

\bibitem{Kazakov:2004nh}
V.~Kazakov and K.~Zarembo, \emph{{Classical / quantum integrability in
  non-compact sector of AdS/CFT}}, {\emph{JHEP} {\bfseries 10} (2004) 060}
  [\href{https://arxiv.org/abs/hep-th/0410105}{{\ttfamily hep-th/0410105}}].

\bibitem{Casteill:2007ct}
P.Y.~Casteill and C.~Kristjansen, \emph{{The Strong Coupling Limit of the
  Scaling Function from the Quantum String Bethe Ansatz}},
  \href{https://doi.org/10.1016/j.nuclphysb.2007.06.011}{\emph{Nucl. Phys.}
  {\bfseries B785} (2007) 1} [\href{https://arxiv.org/abs/0705.0890}{{\ttfamily
  0705.0890}}].

\bibitem{Belitsky:2006en}
A.~Belitsky, A.~Gorsky and G.~Korchemsky, \emph{{Logarithmic scaling in
  gauge/string correspondence}},
  \href{https://doi.org/10.1016/j.nuclphysb.2006.04.030}{\emph{Nucl.Phys.}
  {\bfseries B748} (2006) 24}
  [\href{https://arxiv.org/abs/hep-th/0601112}{{\ttfamily hep-th/0601112}}].

\bibitem{Frolov-Tseytlin-l}
S.~{Frolov} and A.A.~{Tseytlin}, \emph{{Semiclassical quantization of rotating
  superstring in AdS$_{5}{\times}$ S$^{5}$}},
  \href{https://doi.org/10.1088/1126-6708/2002/06/007}{\emph{Journal of High
  Energy Physics} {\bfseries 6} (2002) 7}
  [\href{https://arxiv.org/abs/arXiv:hep-th/0204226}{{\ttfamily
  arXiv:hep-th/0204226}}].

\bibitem{Kostov:1992pn}
I.~Kostov and M.~Staudacher, \emph{{Multicritical phases of the O(n) model on a
  random lattice}},
  \href{https://doi.org/10.1016/0550-3213(92)90576-W}{\emph{Nucl. Phys.}
  {\bfseries B384} (1992) 459}
  [\href{https://arxiv.org/abs/hep-th/9203030}{{\ttfamily hep-th/9203030}}].

\bibitem{Derkachov:2019tzo}
S.~Derkachov and E.~Olivucci, \emph{{Exactly solvable magnet of conformal spins
  in four dimensions}},
  \href{https://doi.org/10.1103/PhysRevLett.125.031603}{\emph{Phys. Rev. Lett.}
  {\bfseries 125} (2020) 031603}
  [\href{https://arxiv.org/abs/1912.07588}{{\ttfamily 1912.07588}}].

\bibitem{Coronado:2018ypq}
F.~Coronado, \emph{{Perturbative four-point functions in planar $ \mathcal{N}=4
  $ SYM from hexagonalization}},
  \href{https://doi.org/10.1007/JHEP01(2019)056}{\emph{JHEP} {\bfseries 01}
  (2019) 056} [\href{https://arxiv.org/abs/hep-th/1811.00467}{{\ttfamily
  hep-th/1811.00467}}].

\bibitem{Kostov:2021omc}
I.~Kostov and V.B.~Petkova, \emph{{Octagon with finite bridge: free fermions
  and determinant identities}},
  \href{https://doi.org/10.1007/JHEP06(2021)098}{\emph{JHEP} {\bfseries 06}
  (2021) 098} [\href{https://arxiv.org/abs/2102.05000}{{\ttfamily
  2102.05000}}].

\bibitem{Arkani-Hamed:2022cqe}
N.~Arkani-Hamed, A.~Hillman and S.~Mizera, \emph{{Feynman polytopes and the
  tropical geometry of UV and IR divergences}},
  \href{https://doi.org/10.1103/PhysRevD.105.125013}{\emph{Phys. Rev. D}
  {\bfseries 105} (2022) 125013}
  [\href{https://arxiv.org/abs/2202.12296}{{\ttfamily 2202.12296}}].

\bibitem{DiFrancesco:1993nw}
P.~Di~Francesco, P.H.~Ginsparg and J.~Zinn-Justin, \emph{{2-D Gravity and
  random matrices}},
  \href{https://doi.org/10.1016/0370-1573(94)00084-G}{\emph{Phys. Rept.}
  {\bfseries 254} (1995) 1}
  [\href{https://arxiv.org/abs/hep-th/9306153}{{\ttfamily hep-th/9306153}}].

\bibitem{BES}
N.~Beisert, B.~Eden and M.~Staudacher, \emph{{Transcendentality and crossing}},
  {\emph{J. Stat. Mech.} {\bfseries 0701} (2007) P021}
  [\href{https://arxiv.org/abs/hep-th/0610251}{{\ttfamily hep-th/0610251}}].

\bibitem{Coronado:2018cxj}
F.~Coronado, \emph{{Bootstrapping the Simplest Correlator in Planar $\mathcal N
  = 4$ Supersymmetric Yang-Mills Theory to All Loops}},
  \href{https://doi.org/10.1103/PhysRevLett.124.171601}{\emph{Phys. Rev. Lett.}
  {\bfseries 124} (2020) 171601}
  [\href{https://arxiv.org/abs/1811.03282}{{\ttfamily 1811.03282}}].

\bibitem{Belitsky:2019fan}
A.~Belitsky and G.~Korchemsky, \emph{{Exact null octagon}},
  \href{https://doi.org/10.1007/JHEP05(2020)070}{\emph{JHEP} {\bfseries 05}
  (2020) 070} [\href{https://arxiv.org/abs/1907.13131}{{\ttfamily
  1907.13131}}].

\bibitem{Belitsky:2020qrm}
A.~Belitsky and G.~Korchemsky, \emph{{Octagon at finite coupling}},
  \href{https://doi.org/10.1007/JHEP07(2020)219}{\emph{JHEP} {\bfseries 07}
  (2020) 219} [\href{https://arxiv.org/abs/2003.01121}{{\ttfamily
  2003.01121}}].

\bibitem{Belitsky:2020qir}
A.V.~Belitsky and G.P.~Korchemsky, \emph{{Crossing bridges with strong
  Szeg\H{o} limit theorem}},
  \href{https://doi.org/10.1007/JHEP04(2021)257}{\emph{JHEP} {\bfseries 04}
  (2021) 257} [\href{https://arxiv.org/abs/2006.01831}{{\ttfamily
  2006.01831}}].

\bibitem{Belitsky:2020qzm}
A.V.~Belitsky, \emph{{Null octagon from Deift-Zhou steepest descent}},
  \href{https://doi.org/10.1016/j.nuclphysb.2022.115844}{\emph{Nucl. Phys. B}
  {\bfseries 980} (2022) 115844}
  [\href{https://arxiv.org/abs/2012.10446}{{\ttfamily 2012.10446}}].

\bibitem{Bargheer:2019kxb}
T.~Bargheer, F.~Coronado and P.~Vieira, \emph{{Octagons I: Combinatorics and
  Non-Planar Resummations}},
  \href{https://doi.org/10.1007/JHEP08(2019)162}{\emph{JHEP} {\bfseries 19}
  (2020) 162} [\href{https://arxiv.org/abs/1904.00965}{{\ttfamily
  1904.00965}}].

\bibitem{Bargheer:2019exp}
T.~Bargheer, F.~Coronado and P.~Vieira, \emph{{Octagons II: Strong Coupling}},
  \href{https://arxiv.org/abs/1909.04077}{{\ttfamily 1909.04077}}.

\bibitem{Olivucci:2021pss}
E.~Olivucci and P.~Vieira, \emph{{Stampedes I: fishnet OPE and octagon
  Bootstrap with nonzero bridges}},
  \href{https://doi.org/10.1007/JHEP07(2022)017}{\emph{JHEP} {\bfseries 07}
  (2022) 017} [\href{https://arxiv.org/abs/2111.12131}{{\ttfamily
  2111.12131}}].

\bibitem{Ahn:2021emp}
C.~Ahn, L.~Corcoran and M.~Staudacher, \emph{{Combinatorial solution of the
  eclectic spin chain}},
  \href{https://doi.org/10.1007/JHEP03(2022)028}{\emph{JHEP} {\bfseries 03}
  (2022) 028} [\href{https://arxiv.org/abs/2112.04506}{{\ttfamily
  2112.04506}}].

\bibitem{Ahn:2022snr}
C.~Ahn and M.~Staudacher, \emph{{Spectrum of the hypereclectic spin chain and
  P\'olya counting}},
  \href{https://doi.org/10.1016/j.physletb.2022.137533}{\emph{Phys. Lett. B}
  {\bfseries 835} (2022) 137533}
  [\href{https://arxiv.org/abs/2207.02885}{{\ttfamily 2207.02885}}].

\bibitem{2005cond.mat..2314A}
D.~{Allison} and N.~{Reshetikhin}, \emph{{Numerical study of the 6-vertex model
  with domain wall boundary conditions}}, {\emph{In Annales de l'institut
  Fourier (Vol. 55, No. 6, pp. 1847-1869)} (2005) }
  [\href{https://arxiv.org/abs/cond-mat/0502314}{{\ttfamily
  cond-mat/0502314}}].

\end{thebibliography}
%  % \bibliographystyle{/Users/vani/Files/PAPERS/PAPERSLIBRARY/utcaps}
% \bibliographystyle{JHEP}

\providecommand{\href}[2]{#2}\begingroup\raggedright\endgroup

 \end{document}